\documentclass[
twocolumn,secnumarabic
,amssymb, amsmath,nobibnotes, aps, prl]{revtex4}
\usepackage[dvips]{graphicx}
\usepackage{tabularx}
\usepackage{bm}%
\expandafter\ifx\csname package@font\endcsname\relax\else
 \expandafter\expandafter
 \expandafter\usepackage
 \expandafter\expandafter
 \expandafter{\csname package@font\endcsname}%
\fi

\begin{document}

\title{The ${\rm AdS}^2_{\theta}/{\rm CFT}_1$ Correspondence and Noncommutative Geometry I:\\
A ${\rm QM}/{\rm NCG}$ Correspondence
}
\author{Badis Ydri}
%\email{ydri@stp.dias.ie, bydri@ictp.it}
\affiliation{Department of Physics, Badji-Mokhtar Annaba University,\\
 Annaba, Algeria.}
%\date{}

\begin{abstract}

A consistent ${\rm QM}/{\rm NCG}$ duality is put forward as a model for the ${\rm AdS}^2/{\rm CFT}_1$ correspondence. This is a duality/correspondence between 1) the dAFF conformal quantum mechanics (${\rm QM}$) on the boundary (which is only "quasi-conformal" in the sense that there is neither an $SO(1,2)-$invariant vacuum state nor there are strictly speaking primary operators), and between 2) the noncommutative geometry   of ${\rm AdS}^2_{\theta}$ in the bulk  (which is only "quasi-AdS" in the sense of being only asymptotically ${\rm AdS}^2$). The Laplacian operators on noncommutative ${\rm AdS}^2_{\theta}$ and commutative ${\rm AdS}^2$ have the same spectrum and thus their correlators are conjectured to be identical. These bulk correlation functions are found to be correctly reproduced by appropriately defined boundary quantum observables in the dAFF quantum mechanics. Moreover, these quasi-primary operators on the boundary form a subalgebra of the operator algebra of  noncommutative ${\rm AdS}^2_{\theta}$.

\end{abstract}

\maketitle
\tableofcontents
\section{Introduction}

In this article a synthesis of the principles of noncommutative geometry and their matrix models together with the principles of the ${\rm AdS}^{2}/{\rm CFT}_1$ correspondence is presented. The main focus will be on constructing a consistent ${\rm QM}/{\rm NCG}$ correspondence, i.e. a correspondence between the dAFF conformal quantum mechanics on the boundary (which is only "quasi-conformal" in the sense of field theory) and the noncommutative geometry   of ${\rm AdS}^2_{\theta}$ in the bulk  (which is only "quasi-AdS" in the sense of being only asymptotically ${\rm AdS}^2$).

The key topic of interest here is noncommutative geometry (thought of as first quantization of geometry) and its Yang-Mills matrix models (which capture among other things the quantum gravitational fluctuations around the noncommutative/matrix backgrounds). This topic finds its root in fuzzy physics (see \cite{Ydri:2001pv} for one of the earliest presentation) and strive to reach a non-perturbative lattice-like matrix-based approach to superstring theory (see \cite{Hanada:2016jok} for a lucid presentation).

\subsection{Symmetric Spaces}
We start off with a brief discussion of the differential geometry of maximally symmetric spaces in two dimensions such as $\mathbb{S}^2$, ${\rm dS}^2$, $\mathbb{H}^2$ and ${\rm AdS}^2$ which play a prominent role in the near-horizon geometry of black holes, in Euclidean quantum field theory, in noncommutative geometry and in the  ${\rm AdS}^{d+1}/{\rm CFT}_d$ correspondence.

First we note that maximally symmetric spaces are essential ingredient in quantum gravity theories and cosmological models. These homogeneous and isotropic spaces enjoy the largest possible amount of spactime symmetries (isometries) and in Lorentzian signature they are exhausted with the three maximally symmetric spaces \cite{Bengtsson}:
\begin{itemize}
\item 1) The de Sitter spacetime ${\rm dS}^d$ (positive scalar curvature, repulsive cosmological constant, topology $\mathbb{S}^{d-1}\times \mathbb{R}$) which is relevant to cosmology.  The de Sitter spacetime ${\rm dS}^d$ as embedded in $\mathbb{M}^{1,d}$ is given by the ambiant metric and the quadric form
\begin{eqnarray}
&&ds^2=-dX_1^2+dX_2^2+...+dX_{d+1}^2\nonumber\\
&&-X_1^2+X_2^2+...+X_{d+1}^2=R^2.
\end{eqnarray}
\item 2) Minkowski spacetime $\mathbb{M}^d$ (zero scalar curvature, zero cosmological constant, topology $\mathbb{R}^d$) which can be viewed as a zero cosmological constant limit of de Sitter spacetime  $\mathbb{dS}^d$.
\item 3) The anti-de Sitter spacetime ${\rm AdS}^d$ (negative scalar curvature, attractive cosmological constant,  topology $\mathbb{R}^{d-1}\times \mathbb{S}^1$) which is relevant to quantum gravity. The anti-de Sitter spacetime ${\rm AdS}^d$ as embedded in $\mathbb{M}^{2,d-1}$ is given by the ambiant metric and the quadric form
\begin{eqnarray}
&&ds^2=-dX_1^2-dX_2^2+dX_3^2+...+dX_{d+1}^2\nonumber\\
&&-X_1^2-X_2^2+X_3^2+...+X_{d+1}^2=-R^2.
\end{eqnarray}
\end{itemize}

However, Wick rotation to Euclidean signature remains crucial to both quantum field theory and noncommutative geometry where quantization of fields and geometries makes strict sense only in Euclidean setting. In Euclidean signature, the maximally symmetric spaces are then given by the three spaces \cite{Bengtsson}:
\begin{itemize}
\item 1) The sphere $\mathbb{S}^d$ (positive curvature). The sphere $\mathbb{S}^d$ as embedded in $\mathbb{R}^{d+1}$ is given by the ambiant metric and the quadric form
\begin{eqnarray}
&&ds^2=dX_1^2+...+dX_{d+1}^2\nonumber\\
&&X_1^2+...+X_{d+1}^2=R^2.
\end{eqnarray}
The Killing vectors fields which leave both the ambiant metric and the quadric form invariant are
\begin{eqnarray}
J_{\alpha\beta}=X_{\alpha}\partial_{\beta}-X_{\beta}\partial_{\alpha}.\label{isometry}
\end{eqnarray}
These $d(d+1)/2$ isometries generate the group of rotations $SO(d+1)$. This is to be contrasted with the isometry group of de Sitter spacetime is $SO(1,d)$.
\item 2) Euclidean space $\mathbb{R}^d$ (zero curvature).
\item 3) The pseudo-sphere $\mathbb{H}^d$ (negative curvature).  The pseudo-sphere (Hyperboloic space) $\mathbb{H}^d$  as embedded in $\mathbb{M}^{1,d}$ is given by the ambiant metric and the quadric form
\begin{eqnarray}
&&ds^2=-dX_1^2+dX_2^2+...+dX_{d+1}^2\nonumber\\
&&-X_1^2+X_2^2+...+X_{d+1}^2=-R^2.
\end{eqnarray}
The Hyperboloic space $\mathbb{H}^d$ is defined as the upper sheet of the two-sheeted hyperboloid $-X_1^2+X_2^2...+X_{d+1}^2=-R^2$. The Killing vectors fields which leave both the ambiant metric and the quadric form invariant are still given by (\ref{isometry}) but the underlying symmetry group is now given by $SO(1,d)$. This is to be contrasted with the isometry group of anti-de Sitter spacetime which is given by $SO(2,d-1)$.

\end{itemize}
It is intriguing to note that in two dimensions the spaces $\mathbb{S}^2$, ${\rm dS}^2$, $\mathbb{H}^2$ and ${\rm AdS}^2$ are simply related. For example, we can go from  ${\rm AdS}^2$ (closed timelike curves with isometry group $SO(2,1)$) to  ${\rm dS}^2$ (closed spacelike curves with isometry group $SO(1,2)$) and vice versa by switching the meaning of timelike and spacelike. While both ${\rm dS}^2$ and $\mathbb{H}^2$ share precisely the same isometry group $SO(1,2)$. And we can go from ${\rm dS}^2$ to $\mathbb{S}^2$ by an ordinary Wick rotation. We can also go from ${\rm AdS}^2$ to $\mathbb{H}^2$ by a Wick rotation.

Representation theory of the Lorentz groups $SO(1,2)$ and $SO(2,1)$ can be found for example in \cite{barg,bns}. See also \cite{Mukunda:1974gb,Girelli:2015ija,Basu:1981ju}.

In this article we will focus on the case of two dimensions with Euclidean signature where the positive curvature space is given by a sphere $\mathbb{S}^2$ with isometry group $SO(3)$ and the negative curvature space is given by a pseudo-sphere $\mathbb{H}^2$  with isometry group $SO(1,2)$. We will be mostly interested in the case of the pseudo-sphere $\mathbb{H}^2$ which we will simply denote by ${\rm AdS}^2$.

The quantization of these two spaces yields the fuzzy sphere $\mathbb{S}^2_N$  \cite{Hoppe,Madore:1991bw} and the noncommutative pseudo-sphere ${\rm AdS}^2_{\theta}$ \cite{Ho:2000fy,Ho:2000br,Jurman:2013ota,Pinzul:2017wch} respectively which enjoy the same isometry groups $SO(3)$ and $SO(1,2)$ as their commutative counterparts. The fuzzy sphere is unstable and suffers collapse in a phase transition to Yang-Mills matrix models (topology change or geometric transition) whereas the noncommutative pseudo-sphere can sustain black hole configurations (by including a dilaton field) and also suffers collapse in the form of the information loss process (quantum gravity transition).

In fact, the product space $\mathbb{S}^2\times {\rm AdS}^2$ is the near-horizon geometry of extremal black holes in general relativity and string theory, e.g. the four-dimensional Reissner-Nordstrom black hole. It is then observed that the information loss problem in  four dimensions on  $\mathbb{S}^2_N\times {\rm AdS}^2_{\theta}$ reduces to the information loss problem in two dimensions on  noncommutative ${\rm AdS}^2_{\theta}$.

As we have said we will be mostly interested here in the case of the pseudo-sphere $\mathbb{H}^2$ or the Euclidean ${\rm AdS}^2$. The goal naturally is to construct a consistent ${\rm AdS}^2$/${\rm CFT}_1$ correspondence.

\subsection{The ${\rm AdS}^{d+1}$/${\rm CFT}_d$ Correspondence} 

Let us first attempt to summarize the original ${\rm AdS}^{d+1}$/${\rm CFT}_d$ correspondence which is the most celebrated example of the gauge/gravity duality and a concrete realization of the holographic principle.

It starts by asserting that maximally supersymmetric  $(p+1)-$dimensional U(N) gauge theory is equivalent to type II superstring theory around black p-brane background spacetime  \cite{Gibbons:1987ps2,Horowitz:1991cd2,Itzhaki:1998dd2} . This is the original conjecture of Maldacena \cite{Maldacena:1997re2} that weakly coupled super Yang-Mills theory and weakly coupled type II superstring theory both provide a description of $N$ coincident Dp-branes \cite{Polchinski:1995mt2} forming a black p-brane. For instance, it is established that the action of maximally supersymmetric gauge theory gives indeed an effective description of Dp-branes in the low energy limit \cite{Witten:1995im2,Dai:1989ua2}.

In the most important cases the near-horizon geometry of these (near-extremal) black p-brane solutions is given by an ${\rm AdS}$ spacetime times a higher dimensional sphere. The  ${\rm AdS}^{d+1}$/${\rm CFT}_d$ correspondence states then that the ${\rm CFT}_d$ generating functional with source $J=\phi_0$ is equal to the path integral on the gravity side evaluated over a bulk field which has the value $\phi_0$ at the boundary of ${\rm AdS}^{d+1}$  \cite{Gubser:1998bc,Witten:1998qj}.

This gauge/gravity duality is then in the words of \cite{Horowitz:2006ct2} the statement that "Hidden within every non-Abelian gauge theory, even within the weak and strong nuclear interactions, is a theory of quantum gravity". 
%Let us then consider a system of $N$ coincident Dp-branes described by the $U(N)$ gauge theory (\ref{run}). This is a theory characterized by the gauge coupling constant $g_{\rm YM}^2\equiv g^2/V_{9-p}$ and the number of colors $N$.

A gauge theory with an infinite number of degrees of freedom, which is the one relevant to supergravity and superstring and which must live in more dimensions by the holographic principle \cite{tHooft:1993dmi2,Susskind:1994vu2} in order to avoid the Weinberg-Witten no-go theorem \cite{Weinberg:1980kq2}, is given by the t'Hooft planar limit \cite{tHooft:1973alw2} in which $N$ (rank of the gauge group) is taken large and $g_{\rm YM}^2$ (gauge coupling constant) is taken small keeping fixed the t'Hooft coupling $\lambda$ given by $\lambda=g_{\rm YM}^2N$.

In summary, the map between the gauge and gravity sides goes as follows:

\begin{itemize}
\item The gauge theory in the limit $N\longrightarrow\infty$ (where extra dimensions will emerge) and $\lambda\longrightarrow\infty$ (where strongly quantum gauge fields give rise to effective classical gravitational fields) should be equivalent to classical type II supergravity around the p-brane spacetime.
\item The gauge theory with $1/N^2$ corrections should correspond to quantum loop corrections, i.e. corrections in $g_s$, in the gravity/string side.
\item The gauge theory with $1/\lambda$ corrections should correspond to stringy corrections, i.e. corrections in $l_s$, corresponding to the fact that degrees of freedom in the gravity/string side are really strings and not point particles.
\end{itemize}
This equivalence should be properly understood as a non-perturbative definition of string theory since the gauge theory is rigorously defined by a lattice \`a la Wilson \cite{Wilson:1974sk2}. For example see \cite{Hanada:2016jok2,OConnor:2016gbq2}.

\subsection{A ${\rm QM}$/${\rm NCG}$ Duality}
In this article we are mostly interested in the case of the pseudo-sphere $\mathbb{H}^2$ which is Euclidean ${\rm AdS}^2$. The goal, as we have said,  is to construct a consistent ${\rm AdS}^2$/${\rm CFT}_1$ correspondence.

The fact that we have for ${\rm AdS}^2$ two disconnected one-dimensional boundaries makes this case very different from higher dimensional anti-de Sitter spacetimes and is probably what makes the ${\rm AdS}^2$/${\rm CFT}_1$ correspondence the most mysterious case among all examples of the AdS/CFT correspondence. For example, see \cite{Strominger:1998yg,Spradlin:1999bn} and \cite{Cadoni:1998sg,Cadoni:1999ja}.

In this article we will instead trace the difficulty of the  ${\rm AdS}^2$/${\rm CFT}_1$ correspondence to the fact that the conformal quantum mechanics residing at the boundary is only quasi-conformal (in a sense to be {\it specified}) and as a consequence the theory in the bulk is only required to be quasi-AdS (in a sense to be {\rm proposed}). In other words, we will take the opposite view and start or assume that the ${\rm CFT}_1$ theory on the boundary is really given by the dAFF conformal quantum mechanics \cite{deAlfaro:1976vlx,Chamon:2011xk} (see also  \cite{Okazaki:2015lpa,Okazaki:2017lpn,Gupta:2019cmo,Gupta:2017lwk,Gupta:2015uga,Gupta:2013ata}).

It is then observed that the Lorentz group $SO(1,2)$ is the fundamental unifying structure of 1) the ${\rm AdS}^2$ spacetime, 2) the noncommutative ${\rm AdS}^2_{\theta}$ space, 3) the geometry of the boundary (which is common to both ${\rm AdS}^2$ and  ${\rm AdS}^2_{\theta}$), 4) the boundary quantum theory, and of 5) the Yang-Mills matrix models defining quantum gravity (or second quantization of the geometry).

In particular, the algebra of quasi-primary operators on the boundary is seen to be a subalgebra of the operator algebra of  noncommutative ${\rm AdS}^2_{\theta}$. This leads us to the conclusion/conjecture that the theory in the bulk must be given by noncommutative geometry \cite{Connes:1996gi} and not by classical gravity, i.e. it is given by ${\rm AdS}^2_{\theta}$ and not by ${\rm AdS}^2$.  We end up therefore with an ${\rm AdS}^2_{\theta}$/${\rm CFT}_1$ correspondence where the ${\rm CFT}_1$ is given by the dAFF conformal quantum mechanics.

In summary, it is observed that the one-dimensional conformal group $SO(1,2)$ is the fundamental unifying structure of the three spaces: 1) The commutative/classical ${\rm AdS}^2$ spacetime, 2) The noncommutative ${\rm AdS}^2_{\theta}$ and 3) The boundary quantum theory. The following set of observations summarizes our main points:
\begin{itemize}
\item The structure of the noncommutative bulk is characterized by full invariance under the group $SO(1,2)$ (bulk isometries are intact and exact).
\item At large distances noncommutative ${\rm AdS}^2_{\theta}$ becomes commutative ${\rm AdS}^2$. The commutative limit defines classical gravity.
\item The near-boundary geometry is commutative even at small distances, i.e. the noncommutative ${\rm AdS}^2_{\theta}$ is asymptotically ${\rm AdS}^2$. The noncommutative  ${\rm AdS}^2_{\theta}$ gravity theory in the bulk is conjectured to be quasi-AdS for all values of the noncommutativity parameter.
\item The quantum mechanical theory at the boundary is an $SO(1,2)-$symmetric quantum mechanics (dAFF conformal quantum mechanics) with quasi-conformal field theoretical structure but with the correct state-operator correspondence.
\item The quasi-primary operators on the boundary define also a spectral triple corresponding to a one-dimensional geometry. The algebra in the bulk reduces to the algebra at the boundary in the stereographical limit.

\item Equivalently, the quasi-conformal structure at the boundary corresponds in fact to the noncommutative ${\rm AdS}^2_{\theta}$ (which is only a quasi-AdS space). So, we also end up with an implicit map between noncommutative  ${\rm AdS}^2_{\theta}$ and commutative  ${\rm AdS}^2$ through the dAFF conformal quantum mechanics on the boundary.  
\item Also, a map between bulk Moyal-Weyl operators and a boundary Heisenberg algebra with an $SO(1,2)-$invariant Hamiltonian is also constructed.
\end{itemize}
This article is organized as follows.

In section $2$ we present a summary of the representation theory of the Lorentz group in three dimensions $SO(1,2)$. Section $3$ contains a detailed construction of the spectral triple corresponding to noncommutative ${\rm AdS}^2_{\theta}$. The commutative limit and the operator definition of the boundary (which is in fact commutative) are discussed in section $4$. The star product and Weyl map of  ${\rm AdS}^2_{\theta}$ are introduced in section $5$ while the corresponding noncommutative actions are introduced in section $6$.  Section $7$ contains our first discussion of the ${\rm AdS}^2_{\theta}/{\rm CFT}_1$ correspondence. In section $8$ we introduce the dAFF conformal quantum mechanics while the operator/state correspondence is introduced in section $9$. The spectral triple of the commutative geometry of the boundary is discussed in section $10$ and a Moyal-Weyl bulk-boundary map is constructed in section $11$. The conclusion contains a summary of the main results/observations of this article and in the appendix we briefly highlight the differences between Lorentzian and Euclidean ${\rm AdS}^2_{\theta}$.

\section{The Lorentz group in three dimensions}
In two dimensions Lorentzian and Euclidean anti-de Sitter spacetimes are simply related. For example, the isometry group of Lorentzian  ${\rm AdS}^2$ is $SO(2,1)$ whereas the isometry group of Euclidean  ${\rm AdS}^2$ is $SO(1,2)$. The Lie algebra for both groups is $su(1,1)$ given explicitly by
\begin{eqnarray}
[K^a,K^b]=if^{ab}~cK^c.\label{so12}
\end{eqnarray}
The structure constants are given by $f^{ab}~c=-\epsilon^{ab}~c$ for Lorentzian ${\rm AdS}^2$ and $f^{ab}~c=\epsilon^{ab}~c$ for Euclidean ${\rm AdS}^2$. This algebra can be put in the canonical form
\begin{eqnarray}
[K^0,K^{\pm}]=\pm K^{\pm}~,~[K^+,K^-]=-2K^0.
\end{eqnarray}
For Lorentzian ${\rm AdS}^2$ we have $K^0=K^3$ and $K^{\pm}=K^1\pm i K^2$ whereas for Euclidean ${\rm AdS}^2$ we have $K^0=K^1$ and $K^{\pm}=-K^3\pm i K^2$.  The Casimir operator is given by
\begin{eqnarray}
C&=&-K_1^2-K_2^2+K_3^2\nonumber\\
&=&K^0(K^0+1)-K^-K^+~,~{\rm Lorentzian}~{\rm AdS}^2.
\end{eqnarray}
\begin{eqnarray}
C&=&-K_1^2+K_2^2+K_3^2\nonumber\\
&=&-K^0(K^0+1)+K^-K^+~,~{\rm Euclidean}~{\rm AdS}^2.
\end{eqnarray}
From this it is quite clear that Euclidean ${\rm AdS}^2$ is obtained from Lorentzian ${\rm AdS}^2$ by the Wick rotation $K_2\longrightarrow -iK_2$.

There are several classes of irreducible representations of $su(1,1)$ calculated originally by Bargmann \cite{barg,bns} which are characterized by a pseudo-spin quantum number $k$. The Hilbert space corresponding to the irreducible representation specified by the number $k$ is given by the equations
\begin{eqnarray} 
  &&K^0|km\rangle =m|km\rangle\nonumber\\
  &&K^+|km\rangle=\sqrt{m(m+1)-k(k-1)}|km+1\rangle\nonumber\\
  &&K^-|km\rangle=\sqrt{m(m-1)-k(k-1)}|km-1\rangle\nonumber\\
   &&C|km\rangle=\pm k(k-1)|km\rangle.
\end{eqnarray}
The plus sign corresponds to Lorentzian  ${\rm AdS}^2$ whereas the minus sign corresponds to Euclidean  ${\rm AdS}^2$.

The possible classes of irreducible representations of $su(1,1)$ are given by the following cases:
\begin{itemize}
\item The discrete series $D^{\pm}_k$ with $k=\{1/2,1,3/2,2,...\}$. We will only consider integer values of "the $su(1,1)$ spin quantum number" $j\equiv k-1$. These are infinite dimensional unitary irreducible representations corresponding to the lowest  and highest weight states, viz
  \begin{itemize}
  \item The lowest-weight representations $D_k^+$ are given by the Hilbert spaces
    \begin{eqnarray}
      {\cal H}_k=\{|km\rangle; m=k,k+1,k+2,...\}.
    \end{eqnarray}
    \item The highest-weight representations $D_k^-$ are given by the Hilbert spaces
    \begin{eqnarray}
      {\cal H}_k=\{|km\rangle; m=-k,-(k+1),-(k+2),...\}.
    \end{eqnarray}
    %The Casimir in these representations is positive given by $C=k(k-1)$.
  \end{itemize}
\item The continuous series $C_k^{\frac{1}{2}}\equiv P_s^{\frac{1}{2}}$ with $k$ a complex number given by $k=\frac{1}{2}+is$ where $s$ is a real number. These are infinite dimensional unitary irreducible representations corresponding to the Hilbert spaces
  \begin{eqnarray}
      {\cal H}_k=\{|km+\frac{1}{2}\rangle; m=0,\pm 1,\pm 2,...\}.
  \end{eqnarray}
  %The Casimir in these representations is negative given by $C=-s^2-\frac{1}{4}<-\frac{1}{4}$.

\item The complementary series $C_k^{0}\equiv P_k^0$ with $k=\frac{1}{2}+is$ as before or with $k$ a real number in the range $[0,1]$. These are infinite dimensional unitary irreducible representations corresponding to the Hilbert spaces
  \begin{eqnarray}
      {\cal H}_k=\{|km\rangle; m=0,\pm 1,\pm 2,...\}.
  \end{eqnarray}
  %The Casimir in these representations is negative in the range $-\frac{1}{4}<C<0$.
\item The finite dimensional irreducible representations $F_k$ of $su(1,1)$ correspond to $k-1\in \mathbb{N}/2$ and they coincide with the irreducible representations  of $su(2)$ with a spin quantum number $j= k-1$. These representations are not unitary.% with a Casimir $C=|k|(|k|+1)$.

\end{itemize}
The Casimir in these representations is always given by $C=k(k-1)=j(j+1)$ (for Lorentzian ${\rm AdS}^2$) or $C=-k(k-1)=-j(j+1)$ for Euclidean  ${\rm AdS}^2$. As we will see, this will have important consequences for deformation quantization. Indeed, quantized Lorentzian  ${\rm AdS}^2$ requires the use of the continuous series whereas quantized Euclidean  ${\rm AdS}^2$ requires the use of the discrete series.
\section{The spectral triple ${\rm AdS}^2_{\theta}$}

The two-dimensional Euclidean anti-de Sitter spacetime ${\rm AdS}^2$ is the co-adjoint orbit $SO(1,2)/SO(2)$ which is a symplectic manifold and thus the canonical quantization of the corresponding Poisson structure, which is given by the inverse of the symplectic form on ${\rm AdS}^2$, produces the noncommutative ${\rm AdS}^2_{\theta}$. See for example  \cite{Ho:2000fy,Ho:2000br,Jurman:2013ota,Pinzul:2017wch}.

The noncommutative geometry of ${\rm AdS}^2_{\theta}$ is defined in terms of a spectral triple $({\cal A}, {\cal H},\Delta)$ consisting of an algebra ${\cal A}$, a Hilbert space ${\cal H}$ and a Laplacian $\Delta$  \cite{Connes:1996gi}.

More explicitly, the noncommutative ${\rm AdS}^2_{\theta}$ is given by the embedding relation and the commutation relations

\begin{eqnarray}
  -\hat{X}_1^2+\hat{X}_2^2+\hat{X}_{3}^2=-R^2.\label{well1}
\end{eqnarray}
%This is a noncommutative space in which the coordinate operators $\hat{X}^a$ are found to satisfy the commutation relations

\begin{eqnarray}
  [\hat{X}^a,\hat{X}^b]=i\kappa \epsilon^{ab}~_c\hat{X}^c.\label{well2}
\end{eqnarray}
The coordinate operators $\hat{X}^a$ which solve (\ref{well1}) and (\ref{well2}) are given explicitly by

\begin{eqnarray}
    \hat{X}^a=\kappa K^a.\label{well3}
  \end{eqnarray}
As we will see shortly these operators define the algebra of operators ${\cal A}$ on  the noncommutative ${\rm AdS}^2_{\theta}$.

The $K^a$ are the generators of the Lie group $SO(1,2)=SU(1,1)/\mathbb{Z}_2$ in the irreducible representations of the Lie algebra $[K^a,K^b]=i\epsilon^{ab}~_cK^c$  given by the discrete series $D_k^{\pm}$ with $k=\{1/2,1,2/3,2,3/2,...\}$.% (we will only consider integer values of $k$). %See \cite{barg} for the detail of the representation theory of $SO(1,2)$.

%These representations $D_k^{\pm}$ are infinite dimensional unitary irreducible representations corresponding to the lowest  and highest weight states given respectively by the Hilbert spaces

% \begin{eqnarray}
%      {\cal H}_k^{\pm}=\{|km\rangle; m=\pm k,\pm (k+1),\pm(k+2),...\}.
%          \end{eqnarray}
% The Casimir operator is given by $C=-K_1^2+K_2^2+K_3^2=-k(k-1).{\bf 1}$ whereas the action of the generators $K^1$ and  $K^{\pm}=-K^{3}\pm i K^2$ is given by

%\begin{eqnarray}
%  &&K^1|km\rangle =m|km\rangle\nonumber\\
%  &&K^{\pm}|km\rangle=\sqrt{m(m\pm 1)-k(k-1)}|km\pm 1\rangle.
%\end{eqnarray}
The generators of the isometry group $SO(1,2)$ acting on the noncommutative ${\rm AdS}^2_{\theta}$ are given by the outer derivations
\begin{eqnarray}
\hat{\cal K}^a(f)=[K_a,f].\label{well4}
\end{eqnarray}
These derivations define natural derivatives (vector fields) on the noncommutative algebra ${\cal A}$ of ${\rm AdS}^2_{\theta}$. Indeed, we have the correct action $\hat{\cal K}^a(\hat{X}_b)=i\epsilon_{ab}~^c\hat{X}_c$ on the coordinate operators $\hat{X}^a$.

The Laplacian operator which fixes the metric structure on  ${\rm AdS}^2_{\theta}$ is then given in terms of the derivations $\hat{\cal K}^a$ by 
\begin{eqnarray}
\hat{\cal K}^2=-\hat{\cal K}_1^2+\hat{\cal K}_2^2+\hat{\cal K}_3^2.\label{Laplacian}
\end{eqnarray}
The derivations $\hat{\cal K}^a$ (since they are commutators) act clearly both on the left and on the right of the algebra ${\cal A}$. This means in particular that the algebra ${\cal A}$ will decompose under the action of the isometry group $SO(1,2)$ as the tensor product of two identical discrete series $D_k^{\pm}$ (see for example \cite{Mukunda:1974gb,Girelli:2015ija,Basu:1981ju}) 
\begin{eqnarray}
D_k^{\pm}\otimes D_{k^{'}}^{\pm}=\bigoplus_{K=k+k^{'}}^{\infty}D_K^{\pm}.\label{pr1}
\end{eqnarray}
\begin{eqnarray}
D_k^{\pm}\otimes D_{k^{'}}^{\mp}=\bigoplus_{K=K_{\rm min}}^{k-k^{'}}D_K^{\pm}\oplus\bigoplus_{K=K_{\rm min}}^{k^{'}-k}D_K^{\mp}\oplus \int_{\mathbb{R}_+}^{\bigoplus}C^{\epsilon}_{\frac{1}{2}+is} ds.\label{pr2}\nonumber\\
\end{eqnarray}
If $k+k^{'}$ is an integer then $K_{\rm min}=\epsilon=0$ whereas if $k+k^{'}$ is half-integer then $K_{\rm min}=\epsilon=1/2$ while the direct sum $\bigoplus_{K=a}^b$ vanishes if $b<a$. Also the $s-$integration is over $\mathbb{R}_+$ because the representations $C_k^{\epsilon}$ and $C_{-k-1}^{\epsilon}$ are isomorphic. 

The above tensor product is the space of operators (which play the role of noncommutative functions) on  ${\rm AdS}^2_{\theta}$. Every irreducible representation in the direct sum corresponds to a different polarization tensor $T_{KM}$ on  ${\rm AdS}^2_{\theta}$.

We can also define integration on the nocommutative  ${\rm AdS}^2_{\theta}$ by means of the invariant scalar product on the algebra ${\cal A}$ given by the trace  
\begin{eqnarray}
(f,g)=Tr f^{\dagger}g~,~f,g\in {\cal A}.\label{sp}
\end{eqnarray}
In summary, the spectral triple $({\cal A},{\cal H}, \Delta)$ corresponding to the nocommutative  ${\rm AdS}^2_{\theta}$ is given explicitly by
\begin{itemize}
\item Algebra:

\begin{eqnarray}
{\cal A}\equiv D_k^{\pm}\otimes D_{k}^{\pm}~,~D_k^{\pm}\otimes D_{k}^{\mp} .
\end{eqnarray}
\item Hilbert space:
\begin{eqnarray}
      {\cal H}\equiv {\cal H}_k^{\pm}.
          \end{eqnarray}
          \item Laplacian:
          \begin{eqnarray}
\hat{\Delta}\equiv \frac{\hat{\cal K}^2}{R^2}.
\end{eqnarray}

\end{itemize}

\section{The commutative limit and the "noncommutative" commutative boundary}
The deformation parameter $\kappa$ in the fundamental commutation relations (\ref{well2}) is not arbitrary. First, $\kappa$ must clearly approach $0$ in the commutative limit. For Euclidean ${\rm AdS}^2_{\theta}$ the Casimir in the discrete and finite representations is negative whereas in the continuous and complementary series the Casimir is positive (the situation is reversed for Lorentzian ${\rm AdS}^2_{\theta}$). This selects the discrete and finite representations. But the finite representation is not unitary which leaves out the discrete series. Thus, by substituting the solution (\ref{well3}) in the constraint (\ref{well1}) the value of the deformation is found to be  quantized in terms of the $su(1,1)$ spin quantum number $j\equiv k-1$ as follows:
\begin{eqnarray}
\frac{R^2}{\kappa^2}=k(k-1).\label{rel1}
\end{eqnarray}
The commutative limit is then defined by
\begin{eqnarray}
\kappa\longrightarrow 0~,~k\longrightarrow \infty.\label{rel2}
\end{eqnarray}
We have seen that noncommutative functions on  ${\rm AdS}^2_{\theta}$ are operators which belong to the algebra ${\cal A}$  which is viewed as the projective module (\ref{pr1}), i.e. as a tensor decomposition under the isometry group $SO(1,2)$. These noncommutative functions form an orthonormal basis with the scalar product (\ref{sp}) and they are in fact $SU(1,1)$ polarization tensors $T_{KM}$ satisfying

\begin{eqnarray}
{\cal K}^2T_{KM}=-K(K-1)T_{KM}~,~{\cal K}^1T_{KM}=MT_{KM}.
\end{eqnarray}
It is not difficult to verify the behavior  \cite{Ho:2000fy,Ho:2000br,Jurman:2013ota}
\begin{eqnarray}
&&T_{KM}=f_{KM}(\hat{X}^1)(\hat{X}^+)^M~,~M\geq 0\nonumber\\
&&T_{KM}=f_{KM}(\hat{X}^1)(\hat{X}^-)^{-M}~,~M\leq 0.\label{polten}
\end{eqnarray}
This is precisely the correct behavior of the commutative functions $Y_{KM}$ on the commutative ${\rm AdS}^2$. In other words, in the commutative limit $k\longrightarrow \infty$ the coordinate operators $\hat{X}^a$ approach the coordinate functions $X^a$. This means in particular that the spectrum of the noncommutative Laplacian $\hat{\Delta}=\hat{\cal K}^2/R^2$ coincides with the spectrum of the commutative Laplacian $\Delta={\cal K}^2/R^2$ with ${\cal K}^2={\cal K}^a{\cal K}_a$ and ${\cal K}_a$ is now given by the pseudo-angular-momentum generator ${\cal K}_a=-i\epsilon_a~^{bc}{X}_b\partial_c$.

In the commutative limit $k\longrightarrow \infty$ the pseudo-sphere $ -{X}_1^2+{X}_2^2+{X}_{3}^2=-R^2$ can be parameterized by local coordinates. For example, in the so-called Poincar\'e patch we have
 \begin{eqnarray}
&&-X_{1}=\frac{z}{2}(1+\frac{R^2+t^2}{z^2})\nonumber\\
  &&-X_2=R\frac{t}{z}\nonumber\\
&&-X_{3}=\frac{z}{2}(1-\frac{R^2-t^2}{z^2}).
\end{eqnarray}
The induced metric is then given by (with $x^0=t$, $x^1=z$)
 \begin{eqnarray}
ds^2&=&g_{\mu\nu}dx^{\mu}dx^{\nu}=(\eta^{ab}\partial_{\mu}X_a\partial_{\nu}X_b)dx^{\mu}dx^{\nu}\nonumber\\
&=&\frac{R^2}{z^2}(dz^2+dt^2).
\end{eqnarray}
In the Poincare patch the isometry generators are given explicitly by 
\begin{eqnarray}
      i{\cal K}^1&=&\frac{1}{2R}\big[(R^2-z^2+t^2)\partial_t+2zt\partial_z\big]\nonumber\\
      i{\cal K}^2&=&-t\partial_t-z\partial_z\nonumber\\
      i{\cal K}^3&=&\frac{1}{2R}\big[(R^2+z^2-t^2)\partial_t-2zt\partial_z\big].\label{sl2R}
    \end{eqnarray}
    The commutative Laplacian $\Delta$ is given in terms of the induced metric $g_{\mu\nu}$ by the usual formula
    \begin{eqnarray}
\Delta&=&-\frac{1}{\sqrt{g}}\partial_{\mu}(\sqrt{g}g^{\mu\nu}\partial_{\nu})=-\frac{z^2}{R^2}(\partial_z^2+\partial_t^2)\nonumber\\
&=&\frac{1}{R^2}(-{\cal K}_1^2+{\cal K}_2^2+{\cal K}_3^2).
    \end{eqnarray}
    Also in the commutative limit the commutators (\ref{well2}) go over to the Poisson brackets
    
    \begin{eqnarray}
\{X^a,X^b\}=\kappa \epsilon^{ab}~c X^x\Leftrightarrow \{t,z\}=\frac{\kappa z^2}{R^2}.
    \end{eqnarray}
    This is actually the basis for the canonical quantization of Euclidean ${\rm AdS}^2$ and it is also the basis for the derivation of the star product between ${\rm AdS}^2$ functions. This results indicates also explicitly that the coordinate operators $\hat{X}^a$ approach in the commutative limit the coordinate functions $X^a$.

Euclidean  ${\rm AdS}^2$ is obtained from Lorentzian  ${\rm AdS}^2$ by means of the Euclidean continuation $X_2\longrightarrow -iX_2$ which corresponds in the Poincar\'e patch to the Wick rotation $t\longrightarrow -it$.  Hence, the matrix Euclidean continuation $\hat{X}_2\longrightarrow -i\hat{X}_2$ will produce the nocommutative Euclidean  ${\rm AdS}^2_{\theta}$ from the noncommutative Lorentzian  ${\rm AdS}^2_{\theta}$. We also note that in the Minkowski case the Poincar\'e coordinates are called a patch because they cover only half of the spacetime which obviously can be extended beyond the horizon (the point where the Poincar\'e coordinates terminate).

We can think of $z$ as a lattice spacing and $u=1/z$ as an ultraviolet cutoff of the conformal quantum mechanics which lives on the line foliated at the point $z$. The point $z=0$ or equivalently $u=\infty$ is the actual conformal boundary (where the actual conformal quantum mechanics is sustained) whereas the point $z=\infty$ or equivalently $u=0$ is the horizon which is a single point in the Euclidean case.

The radius operator of the  noncommutative Euclidean  ${\rm AdS}^2_{\theta}$ is obviously defined by 

\begin{eqnarray}
      &&\hat{u}=\frac{1}{\hat{z}}=\frac{\hat{X}_1-\hat{X}_3}{R^2}.
    \end{eqnarray}
We compute immediately the expectation value $\langle km|\hat{u}|km\rangle=\kappa m/R^2$ which approaches $\pm \infty$  as $m\longrightarrow \pm\infty$ (corresponding to the two boundaries of noncommutative ${\rm AdS}^2_{\theta}$ represented by the two representations $D_k^{\pm}$). Indeed, in the discrete representations $D_k^+$ and $D_k^-$ we can explicitly calculate the  eigenvalues of the radius operator $\hat{u}$ and find them to be given by the commutative values $u\geq 0$ and $u\leq 0$ respectively. Hence, near the boundaries we have very large values of $m$ \cite{Pinzul:2017wch}.

We can furthermore show that the near-boundary action of a scalar field theory on  noncommutative ${\rm AdS}^2_{\theta}$ is an ordinary free scalar field theory on commutative ${\rm AdS}^2$ (with rescaled fields). This means in particular that noncommutative ${\rm AdS}^2_{\theta}$ is asymptotically a commutative ${\rm AdS}^2$ \cite{Pinzul:2017wch}. This will be discussed further below.

These results show that the distinguished physical observers living at the boundary $z=0$ (which observe quantum mechanics at the boundary and classical general relativity in the bulk) are actually commutative observers unaffected by the quantization of the spactime manifold at the Planck scale.

 %\section{The semi-classical/commutative limit and de-quantization}
 \section{Comparison between the fuzzy sphere $\mathbb{S}^2_N$ and the noncommutative pseudo-sphere $\mathbb{H}^2_{\theta}$}
From an operational point of view the noncommutative Euclidean ${\rm AdS}^2_{\theta}$ (or equivalently the noncommutative pseudo-sphere $\mathbb{H}^2_{\theta}$) is given in terms of the generators $K_a$ of $SO(1,2)$ in the pseudo-spin $k$ irreducible representation $D_k^{\pm}$ by the coordinate operators (\ref{well3}) which satisfy the embedding relation (\ref{well1}) and the commutator eqaution (\ref{well2}) such that the quantization condition (\ref{rel1}) for the deformation parameter $\kappa$ must hold. Namely, we have
\begin{eqnarray}
&&    \hat{X}^a=\kappa K^a\nonumber\\
&&  -\hat{X}_1^2+\hat{X}_2^2+\hat{X}_{3}^2=-R^2\nonumber\\
&&  [\hat{X}^a,\hat{X}^b]=i\kappa \epsilon^{ab}~_c\hat{X}^c\nonumber\\
&& \frac{R^2}{\kappa^2}=k(k-1).
\end{eqnarray}
Let us write for comparison the corresponding equations defining the fuzzy sphere $\mathbb{S}^2_N$. First, let $L_a$ be the generators of $SO(3)$ in the spin $l=(N-1)/2$ irreducible representation,  $r$ be the radius of the sphere and $\alpha$ be the deformation parameter. Then the fuzzy sphere  $\mathbb{S}^2_N$ is defined operationally by the equations
\begin{eqnarray}
&&    \hat{x}^a=\alpha L_a\nonumber\\
&&  \hat{x}_1^2+\hat{x}_2^2+\hat{x}_{3}^2=r^2\nonumber\\
&&  [\hat{x}_a,\hat{x}_b]=i\alpha \epsilon_{abc}\hat{x}_c\nonumber\\
&& \frac{r^2}{\kappa^2}=l(l+1).
\end{eqnarray}
The commutative limit of the sphere is given by $l\longrightarrow\infty$ and of the pseudo-sphere is given by $k\longrightarrow \infty$. In these limits the coordinate operators $\hat{x}_a$ and $\hat{X}_a$ tend to the coordinate functions $x_a$ and $X_a$ respectively and the embedding relations  reduce to
\begin{eqnarray}
{x}_1^2+{x}_2^2+{x}_3^2=r^2~,~\mathbb{S}^2.
\end{eqnarray}
\begin{eqnarray}
&&-{X}_1^2+{X}_2^2+{X}_3^2=-R^2~,~\mathbb{H}^2\nonumber\\
&&-{X}_1^2-{X}_2^2+{X}_3^2=-R^2~,~{\rm AdS}^2.
\end{eqnarray}
On the other hand, the commutator equations become the Poisson brackets 
\begin{eqnarray}
\{{x}_a,{x}_b\}=\alpha\epsilon_{abc}{x}_c.\label{brac1}
\end{eqnarray}
\begin{eqnarray}
\{{X}_a,{X}_b\}=\kappa f_{abc}{X}^c.\label{brac2}
\end{eqnarray}
In the commutative limits  $l\longrightarrow\infty$ and $k\longrightarrow \infty$ we can also work with local coordinates. On the sphere $\mathbb{S}^2$ we can work with the spherical coordinates $\theta$ and $\phi$ given by
\begin{eqnarray}
  &&X_{1}=R\sin\theta\cos\phi~,~X_2=R\sin\theta\sin\phi~,~X_3=R\cos\theta.\nonumber\\
\end{eqnarray}
The coordinates used for the anti-de Sitter spacetime ${\rm AdS}^2$ in the previous section, and in most of this article, are the Poincar\'e coordinates $(t,z)$. A better system of coordinates for  ${\rm AdS}^2$ which covers the whole of spacetime are the cylindrical coordinates $(\tau,\sigma)$ defined by
\begin{eqnarray}
  &&X_{1}=R\frac{\cos \tau}{\cos \sigma}~,~X_2=R\frac{\sin \tau}{\cos \sigma}~,~X_3=R\tan \sigma.
\end{eqnarray}
For the Euclidean anti-de Sitter space, i.e. the pseudo-sphere $\mathbb{H}^2$ the cylindrical coordinates are defined by 
\begin{eqnarray}
   &&X_{1}=R\frac{\cosh \tau}{\cos \sigma}~,~X_2=R\frac{\sinh \tau}{\cos \sigma}~,~X_3=R\tan \sigma.\nonumber\\
\end{eqnarray}
Clearly,  ${\rm AdS}^2$ spacetime and the pseudo-sphere $\mathbb{H}^2$ are connected by a Wick rotation.

The Poisson brackets (\ref{brac1}) and (\ref{brac2}) can also be expressed in local coordinates. Indeed, the Poisson structures on the sphere and on the pseudo-sphere are given respectively by \footnote{The Poisson structure, the corresponding non-degenerate symplectic form and their quantization makes strict sense only in Euclidean signature. Indeed, since the symplectic form is non-degenerate it can be inverted resulting in Poisson structures which can then be quantized in the usual way by replacing them with commutators.}

\begin{eqnarray}
\{f,g\}=\frac{\kappa}{R\sin\theta}(\partial_{\theta}f\partial_{\phi}g-\partial_{\phi}f\partial_{\theta}g)~,~\mathbb{S}^2.
\end{eqnarray}
\begin{eqnarray}
\{f,g\}=\kappa\frac{\cos^2\sigma}{R}(\partial_{\tau}f\partial_{\sigma}g-\partial_{\sigma}f\partial_{\tau}g)~,~\mathbb{H}^2.
\end{eqnarray}
These Poisson structures correspond to non-degenerate symplectic two-forms. For example, the sphere $\mathbb{S}^2$ which is the co-adjoint orbit $SO(3)/U(1)$ admits a non-degenerate symplectic two-form $\omega$ given by 
\begin{eqnarray}
\omega=\frac{R}{\alpha}\sin\theta d\theta\wedge d\phi.%=\frac{\kappa}{2}\epsilon_{abc}X_a dX_b\wedge dX_c.
\end{eqnarray}
Similarly, the pseudo-sphere  $\mathbb{H}^2$  is the co-adjoint orbit $SO(1,2)/U(1)$ which admits a  non-degenerate symplectic two-form $\omega$ given by
\begin{eqnarray}
  \omega=\frac{1}{\kappa}\frac{R}{\cos^2\sigma} d\tau\wedge d\sigma.
\end{eqnarray}
These forms are closed, i.e. $d\omega=0$ and their quantization (which leads to the quantization of the deformation parameters $\alpha$ and $\kappa$) yields the fuzzy sphere $\mathbb{S}^2_N$ and the noncommutative pseudo-sphere $\mathbb{H}^2_{\theta}$ respectively. Alternatively, since these forms are non-degenerate they can be inverted to define Poisson structures which can then be quantized in the usual way by replacing the Poisson brackets with commutators.

The generators of the isometry group $SO(3)$ of the sphere $\mathbb{S}^2$  are given by the angular momentum generators ${\cal L}_a=-i\epsilon_{abc}x_b{\partial}_c$ which acts on the coordinate functions as  ${\cal L}_a(x_b)=-i\epsilon_{abc}x_c$. These are the Killing vector fields generating rotations on the sphere and satisfying the $su(2)$ Lie algebra $[L_a,L_b]=i\epsilon_{abc}L_c$.

Similarly, the generators of the isometry group $SO(1,2)$ of the pseudo-sphere $\mathbb{H}^2$  are given by the "pseudo angular momentum" generators ${\cal K}_a=-if_{ab}~^{c}X^b{\partial}_c$ (here $\partial_c=\partial/\partial X^c$) which acts on the coordinate functions as  ${\cal K}_a(X_b)=-if_{ab}~^{c}X_c$. These are the Killing vector fields generating isometries or pseudo-rotations (translations, dilatations and special conformal transformations) on the pseudo-sphere and satisfying the $su(1,1)$ Lie algebra $[K_a,K_b]=if_{ab}~^cK_c$, i.e. the Lie algebra (\ref{so12}).

The Killing vector fields ${\cal L}_a$ and ${\cal K}_a$ act by outer derivations (commutators) in the noncommutative spaces $\mathbb{S}^2_N$, $\mathbb{H}^2$ and ${\rm AdS}^2_{\theta}$. See for example equation (\ref{well4}).

The Killing vector fields can also be defined in local coordinates. We take as an example the case of anti-de Sitter spacetime and the case of the pseudo-sphere.

The generators of the isometry groups $SO(2,1)$ of ${\rm AdS}^2$ can be represented in terms of the global coordinates $X_A$ as follows
\begin{eqnarray}
L^A_{B}=X^A\frac{\partial}{\partial X^B}-X_B\frac{\partial}{\partial X_A}.
\end{eqnarray}
Explicitly, we have 

\begin{eqnarray}
-i{\cal K}^3\equiv L_1^2&=&X^2\frac{\partial}{\partial X^1}-X_1\frac{\partial}{\partial X_2}=-\partial_{\tau}.
\end{eqnarray}

\begin{eqnarray}
   -i{\cal K}^2\equiv L_3^1&=&X^1\frac{\partial}{\partial X^3}-X_3\frac{\partial}{\partial X_1}\nonumber\\
   &=&\sin\tau\sin\sigma\partial_{\tau}-\cos\tau\cos\sigma\partial_{\sigma}.
\end{eqnarray}
\begin{eqnarray}
  i{\cal K}^1\equiv L_2^3&=&X^3\frac{\partial}{\partial X^2}-X_2\frac{\partial}{\partial X_3}\nonumber\\
  &=&-\cos\tau\sin\sigma\partial_{\tau}-\sin\tau\cos\sigma\partial_{\sigma}.
\end{eqnarray}
%\begin{eqnarray}
%-iK^3\equiv L_2^1=X^1\frac{\partial}{\partial X^2}-X_2\frac{\partial}{\partial X_1}=\cosh\tau\sin \sigma \partial_{\tau}-\sinh\tau \cos \sigma\partial_{\sigma}.
%\end{eqnarray}
%These are the generators of the conformal group on the boundary of ${\rm AdS}^2$.
These generators satisfy the algebra (\ref{so12}) which can be given explicitly by
%\begin{eqnarray}
%[K^1,K^2]=iK^3~,~[K^2,K^3]=-iK^1~,~[K^3,K^1]=iK^2.
%\end{eqnarray}
\begin{eqnarray}
[{\cal K}^3,{\cal K}^2]=-i{\cal K}^1~,~[{\cal K}^1,{\cal K}^2]=-i{\cal K}^3~,~[{\cal K}^3,{\cal K}^1]=i{\cal K}^2.\nonumber\\
\end{eqnarray}
%\begin{eqnarray}
%[K^3,K^2]=iK^1~,~[K^1,K^2]=iK^3~,~[K^3,K^1]=-iK^2.
%\end{eqnarray}
%We write these equations collectively as
%\begin{eqnarray}
%  [K^a,K^b]=i\epsilon^{ab}~_cK^c~,~{\rm Euclidean}.
%\end{eqnarray}
%\begin{eqnarray}
%  [K^a,K^b]=if^{ab}~_cK^c.\label{su11}
%\end{eqnarray}
Similar calculation can be done for the Euclidean anti-de Sitter space. Indeed, the generators of the isometry group $SO(1,2)$ of the pseudo-sphere $\mathbb{H}^2$ are given in the local cylindrical coordinates $(\tau,\sigma)$ by

\begin{eqnarray}
-i{\cal K}^3\equiv L_1^2=X^2\frac{\partial}{\partial X^1}-X_1\frac{\partial}{\partial X_2}=-\partial_{\tau}.
\end{eqnarray}
 \begin{eqnarray}
   i{\cal K}^2\equiv L_3^1&=&X^1\frac{\partial}{\partial X^3}-X_3\frac{\partial}{\partial X_1}\nonumber\\
   &=&\sinh\tau\sin \sigma \partial_{\tau}-\cosh\tau \cos \sigma\partial_{\sigma}.
\end{eqnarray}
\begin{eqnarray}
  -i{\cal K}^1\equiv L_2^3&=&X^3\frac{\partial}{\partial X^2}-X_2\frac{\partial}{\partial X_3}\nonumber\\
  &=&\cosh\tau\sin \sigma \partial_{\tau}-\sinh\tau \cos \sigma\partial_{\sigma}.
\end{eqnarray}
Again we find the algebra (\ref{so12}) which in the Euclidean case reads explicitly as follows

\begin{eqnarray}
[{\cal K}^3,{\cal K}^2]=i{\cal K}^1~,~[{\cal K}^1,{\cal K}^2]=i{\cal K}^3~,~[{\cal K}^3,{\cal K}^1]=i{\cal K}^2.
\end{eqnarray}

\section{Moyal-Weyl star product and Weyl map}

The goal in this section is to write down the star product on noncommutative ${\rm AdS}^2_{\theta}$ using  Darboux theorem which plays in noncommutative geometry and emergent gravity (quantum gravity as defined by matrix models) the same role played by the equivalence principle in differential geometry and classical general relativity. Darboux theorem states that every quantized symplectic manifold is locally a noncommutative Moyal-Weyl space, i.e. there exists a coordinate system $\bar{x}$ in which the noncommutativity parameter $\bar{\theta}^{ij}$ is constant. The role of Darboux theorem in noncommutative geometry and matrix models is discussed for example in \cite{Lee:2010zf1} and \cite{Blaschke:2010ye}.

Again we consider Euclidean ${\rm AdS}^2$ which is a pseudosphere in Minkowski spacetime  with metric $\eta=(-1,+1,+1)$. The Poincare coordinates $(t,z)$ are related to the canonical coordinates $(x,y)$ (in which the noncommutativity parameter is constant \cite{Pinzul:2017wch,deAlmeida:2019awj}) by the following relations

  \begin{eqnarray}
     z=R\exp(-x)~,~t=\exp(-x)y.
    \end{eqnarray}
The metric becomes
 \begin{eqnarray}
     ds^2&=&\frac{R^2}{z^2}(dz^2+dt^2)\nonumber\\
&=&R^2dx^2+(dy-ydx)^2.\label{metric}
    \end{eqnarray}
The boundary  $z\longrightarrow 0$ corresponds to both $x,y\longrightarrow \infty$ in such a way that the combination $t$ is kept fixed.

The explicit relation between the global  embedding coordinates $X_a$ and the local coordinates $(x,y)$ is given by

\begin{eqnarray}
X_3&=&\frac{1}{2R}e^{-x}y^2+\frac{R}{2}e^{-x}-\frac{R}{2}e^x\nonumber\\
X_2&=&y\nonumber\\
X_1&=&\frac{1}{2R}e^{-x}y^2+\frac{R}{2}e^{-x}+\frac{R}{2}e^x.\label{co}
 \end{eqnarray}
The fundamental Poisson bracket in the canonical coordinates $(x,y)$ is then given by
\begin{eqnarray}
\{x,y\}=\kappa.
 \end{eqnarray}
Furthermore, the generators of the $SO(1,2)$ isometry group of Euclidean ${\rm AdS}^{2}$ are given by

 \begin{eqnarray}
      i{\cal K}^1&=&-\frac{1}{R}e^{-x}y\partial_x-X_3\partial_y\nonumber\\
      i{\cal K}^2&=&\partial_x\nonumber\\
      i{\cal K}^3&=&\frac{1}{R}e^{-x}y\partial_x+X_1\partial_y.\label{Ks}
    \end{eqnarray}
After quantization, the operators $\hat{x}$ and $\hat{y}$ associated with the canonical coordinates $x$ and $y$ must therefore satisfy the canonical commutation relation

\begin{eqnarray}
[\hat{x},\hat{y}]=i\kappa.\label{MW}
\end{eqnarray}
This is a Heisenberg structure which allows us to introduce a Weyl map \cite{weyl} and a Moyal-Weyl star product \cite{Moyal:1949skv2,Groenewold:1946kpv2} in the usual way.

The weyl map $\pi$ allows us to map operators $\hat{F}(\hat{x},\hat{y})$ in the Hilbert space ${\cal H}_k^{\pm}$ back to functions $\pi(\hat{F})(x,y)$ on commutative ${\rm AdS}^2$ such that the operator product $\hat{F}.\hat{G}$ of two operators $\hat{F}$ and $\hat{G}$ is mapped to the star product $\pi(\hat{F})*{\pi}(\hat{G})$ of the two corresponding functions $\pi(\hat{F})$ and $\hat{\pi}(\hat{G})$, namely

\begin{eqnarray}
\pi(\hat{F}.\hat{G}(\hat{x},\hat{y}))=\pi(\hat{F})*\hat{\pi}(\hat{G})(x,y).
\end{eqnarray}
By construction the map of the canonical operators $\hat{x}$ and $\hat{y}$ is precisely the canonical coordinates $x$ and $y$, viz

\begin{eqnarray}
\pi(\hat{x})=x~,~\pi(\hat{y})=y.
\end{eqnarray}
The star product is given explicitly by  (with $(\bar{x}^1,\bar{x}^2)=(x,y)$ and $\bar{\theta}^{ab}=\kappa \epsilon^{ab}$)

\begin{eqnarray}
  f\bar{*}g(\bar{x})&=&\exp(\frac{i}{2}\bar{\theta}^{ab}\frac{\partial}{\partial\xi^a}\frac{\partial}{\partial\eta^b})f(\bar{x}+\xi)g(\bar{x}+\eta)|_{\xi=\eta=0}.\nonumber\\
\end{eqnarray}
Or equivalently

\begin{eqnarray}
  f\bar{*}g(\bar{x})&=&f({x},y)\exp\big(\frac{i\kappa}{2}(\overleftarrow{{\partial}}_x\overrightarrow{{\partial}}_y-\overleftarrow{{\partial}}_y\overrightarrow{{\partial}}_x)\big)g({x},y).\nonumber\\
\end{eqnarray}
In particular, we have for $f=f(x)$ the results

\begin{eqnarray}
  f\bar{*}g(\bar{x})&=&f(x+\frac{i\kappa}{2}\overrightarrow{{\partial}}_y)g(x,y)\nonumber\\
  g\bar{*}f(\bar{x})&=&g(x,y)f(x-\frac{i\kappa}{2}\overleftarrow{{\partial}}_y).\label{star1}
\end{eqnarray}
Similarly, we have for $g=g(y)$ the results

\begin{eqnarray}
  g\bar{*}f(\bar{x})&=&g(y-\frac{i\kappa}{2}\overrightarrow{{\partial}}_x)f(x,y)\nonumber\\
  f\bar{*}g(\bar{x})&=&f(x,y)g(y+\frac{i\kappa}{2}\overleftarrow{{\partial}}_x).\label{star2}
\end{eqnarray}
The relation between the embedding coordinate operators $\hat{X}_a$ and the canonical coordinate operators $\hat{x}$ and $\hat{y}$ is given from (\ref{co}) by the equations

\begin{eqnarray}
\hat{X}_3&=&\frac{1}{2R}\hat{y}e^{-\hat{x}}\hat{y}+\frac{R}{2}e^{-\hat{x}}-\frac{R}{2}e^{\hat{x}}\nonumber\\
\hat{X}_2&=&\hat{y}\nonumber\\
\hat{X}_1&=&\frac{1}{2R}\hat{y}e^{-\hat{x}}\hat{y}+\frac{R}{2}e^{-\hat{x}}+\frac{R}{2}e^{\hat{x}}.\label{ord}
 \end{eqnarray}
It is then straightforward to check  (using $[\exp(\pm \hat{x}),\hat{y}]=\pm i\kappa\exp(\pm \hat{x})$) that

\begin{eqnarray}
[\hat{X}^3,\hat{X}^2]=i\kappa\hat{X}^1~,~[\hat{X}^1,\hat{X}^2]=i\kappa\hat{X}^3~,~[\hat{X}^3,\hat{X}^1]=i\kappa\hat{X}^2.\nonumber\\
 \end{eqnarray}
These are precisely equations (\ref{well2}). In the same manner we can verify equation (\ref{well1}). This shows explicitly that the operator ordering chosen in (\ref{ord}) is the correct one.

The derivation operators $\hat{\partial}_x$ and $\hat{\partial}_y$ associated with the canonical operators $\hat{x}$ and $\hat{y}$ are given immediately (since (\ref{MW}) defines a Moyal-Weyl plane) by

\begin{eqnarray}
\hat{\partial}_x=-\frac{1}{i\kappa}[\hat{y},]~,~\hat{\partial}_y=\frac{1}{i\kappa}[\hat{x},].
\end{eqnarray}
The noncommutative derivation operator $\hat{\cal K}^2$ associated with the coordinate operator $\hat{X}_2=\hat{y}$ is then given by

\begin{eqnarray}
i\hat{\cal K}^2=\frac{i}{\kappa}[\hat{X}_2,]=\frac{i}{\kappa}[\hat{y},]=\hat{\partial}_x.
\end{eqnarray}
By construction this has the correct commutative limit $i\hat{\cal K}^2\longrightarrow i{\cal K}^2=\partial_x$. Similarly, the noncommutative derivation operators $\hat{\cal K}^1$ and $\hat{\cal K}^3$ should be defined by the relations

\begin{eqnarray}
i\hat{\cal K}^1=\frac{i}{\kappa}[\hat{X}_1,]~,~i\hat{\cal K}^3=\frac{i}{\kappa}[\hat{X}_3,].
\end{eqnarray}
We have explicitly (where the arrow indicates the Weyl map)
\begin{eqnarray}
i\hat{\cal K}^{1,3}(\hat{f})&=&\frac{i}{\kappa}[\hat{X}_{1,3},\hat{f}]\nonumber\\
&=& \frac{i}{\kappa}\bigg(\frac{1}{2R}[\hat{y}e^{-\hat{x}}\hat{y},\hat{f}]+\frac{R}{2}[e^{-\hat{x}},\hat{f}]\pm \frac{R}{2}[e^{\hat{x}},\hat{f}]\bigg).\nonumber\\
\end{eqnarray}
In terms of the Weyl map and the Moyal-Weyl star product this reads
\begin{eqnarray}
i\hat{\cal K}^{1,3}(\hat{f})&\longrightarrow &\frac{i}{\kappa}\bigg(\frac{1}{2R}[{y}*e^{-{x}}*{y},{f}]_*+\frac{R}{2}[e^{-{x}},{f}]_*\nonumber\\
&\pm& \frac{R}{2}[e^{{x}},{f}]_*\bigg).
\end{eqnarray}
It is not difficult to check explicitly (by using (\ref{star1}) and (\ref{star2})) that this definition of $\hat{\cal K}^1$ and $\hat{\cal K}^3$ will tend in the commutative limit $\kappa\longrightarrow 0$ to the generators ${\cal K}^1$ and ${\cal K}^3$ given explicitly in equation (\ref{Ks}). In other words, the generators $\hat{\cal K}^a$ are the noncommutative  ${\rm AdS}^2_{\theta}$ Killing vectors  in the same way that the generators ${\cal K}^a$ are the commutative ${\rm AdS}^2$ Killing vectors.

\section{Noncommutative (scalar) actions}
We are now in a position to write down an action principle for scalar fields $\hat{\Phi}$ on  noncommutative  ${\rm AdS}^2_{\theta}$ using the Killing vectors $\hat{\cal K}^a$ as outer derivations. See in particular \cite{Pinzul:2017wch}. First we note the dimension of the various objects: $[t],[z]\sim R$, $[x]\sim 1$, $[y]\sim R$, $[\kappa]\sim R$ and $[X^a]\sim R$. In particualr note that $x$ is dimensionless and thus the noncommutativity parameter $\bar{\theta}^{ab}\equiv \kappa\epsilon^{ab}$ is of dimension length and not of dimension area. The scalar field $\hat{\Phi}$ in two dimension is canonically of dimension $0$. We can then write down immediately the free action

\begin{eqnarray}
S=\frac{2\pi R \kappa}{2}{\rm Tr}\bigg(m^2\hat{\Phi}^2-\frac{1}{R^2\kappa^2}[\hat{X}^a,\hat{\Phi}][\hat{X}_a,\hat{\Phi}]\bigg).\label{action}
 \end{eqnarray}
The Weyl map takes the trace ${\rm Tr}$ to the integral on ${\rm AdS}^2$ as follows (the metric $g$ in the canonical coordinates is read from the second line of (\ref{metric}) and it is clear that $\sqrt{{\rm det}g}=1$)

\begin{eqnarray}
2\pi {\rm Tr}\longrightarrow \int \frac{dx dy}{\sqrt{{\rm det}\bar{\theta}}}=\int \frac{dx dy}{\kappa}.
 \end{eqnarray}
Recall that the integral over a symplectic manifold ${\cal M}^{d}$, such as    ${\rm AdS}^2$, is given in terms of its symplectic structure $\omega$, which is here $\omega=-dx\wedge dy/\kappa$, by the formula (where $d=2n$)

\begin{eqnarray}
(2\pi)^n{\rm Tr} \longrightarrow \int \frac{\omega^n}{n!}=\int \frac{d^{2n}x}{\sqrt{{\rm det}\bar{\theta}}}.
\end{eqnarray}
In terms of the star product the above action reads then (where ${\cal X}^a=\pi(\hat{X}^a)$)

\begin{eqnarray}
S&=&\frac{R}{2}\int dx dy\bigg(m^2\Phi*\Phi-\frac{1}{R^2\kappa^2}[{\cal X}^a,\Phi]_**[{\cal X}_a,\Phi]_*\bigg).\label{actionstar}\nonumber\\
\end{eqnarray}
Recall that ${\rm AdS}^2$ is a manifold with two boundaries in which boundary terms play an important role. However, for simplicity we start by neglecting boundary terms. Thus, the integral over the star product of two functions $f$ and $g$ is equal (up to boundary terms)  to the integral over the ordinary product of these  two functions, viz

\begin{eqnarray}
\int dx dy f*g(x,y)&=&\int dx dy f(x,y).g(x,y)\nonumber\\
&+&{\rm boundary~ terms}.
\end{eqnarray}
By concentrating on the bulk fields for the moment we can use this equation to obtain the action

\begin{eqnarray}
S
 &=&\frac{1}{2R\kappa^2}\int dx dy\bigg(R^2\kappa^2m^2\Phi^2+[{\cal X}^1,\Phi]_*^2-[{\cal X}^2,\Phi]_*^2\nonumber\\
 &-&[{\cal X}^3,\Phi]_*^2\bigg)\nonumber\\
&=& \frac{1}{2R\kappa^2}\int dx dy\bigg(R^2\kappa^2m^2\Phi^2+[y*e^{-x}*y,\Phi]_*[e^x,\Phi]_*\nonumber\\
&+&R^2[e^{-x},\Phi]_*[e^{x},\Phi]_*-[y,\Phi]_*^2\bigg).
\end{eqnarray}
A relatively short calculation using (\ref{star1}) and (\ref{star2}) gives the results

\begin{eqnarray}
[y,\Phi]_*=-i\kappa\partial_x \Phi.
\end{eqnarray}

\begin{eqnarray}
[e^{\pm x},\Phi]_*=\pm i\kappa e^{\pm x}\Delta_y \Phi.
\end{eqnarray}

\begin{eqnarray}
[y*e^{-x}*y,\Phi]_*&=&-i\kappa e^{-x}\bigg[2yS_y\partial_x\Phi -\frac{\kappa^2}{4}\partial_x^2\Delta_y\Phi\nonumber\\
&+&y^2\Delta_y\Phi+\frac{\kappa^2}{4}\Delta_y\Phi\bigg].
\end{eqnarray}
In the last two equations the operators $\Delta_y$ and $S_y$ are given explicitly by

\begin{eqnarray}
\Delta_y=\frac{2}{\kappa}\sin \frac{\kappa}{2}\partial_y~,~S_y=\cos \frac{\kappa}{2}\partial_y.
\end{eqnarray}
We get then the action

\begin{eqnarray}
S
&=& \frac{1}{2R}\int \bigg(R^2m^2\Phi^2+(\partial_x\Phi)^2+(R^2+\frac{\kappa^2}{4}+y^2)(\Delta_y\Phi)^2\nonumber\\
&+&2y\Delta_y\Phi S_y\partial_x\phi -\frac{\kappa^2}{4}\Delta_y\Phi\partial_x^2\Delta_y\Phi\bigg).
\end{eqnarray}
We use now the identity $1=S_y²+\kappa^2 \Delta_y^2/4$ or equivalently

\begin{eqnarray}
\int  (\partial_x\Phi)^2&=&\int \partial_x\Phi S_y^2\partial_x\Phi+\frac{\kappa^2}{4}\int  \partial_x\Phi\Delta_y^2\partial_x\Phi\nonumber\\
&=& \int  (S_y\partial_x\Phi)^2-\frac{\kappa^2}{4}\int  (\Delta_y\partial_x\Phi)^2.\nonumber\\
\end{eqnarray}
In the last line we have neglected boundary terms and used the results

\begin{eqnarray}
\int f(x,y).\Delta_yg(x,y)&=&-\int  \Delta_yf(x,y).g(x,y)+{\rm b.t}.\nonumber\\
\end{eqnarray}
And
\begin{eqnarray}
\int  f(x,y).S_yg(x,y)&=&\int  S_yf(x,y).g(x,y)+{\rm b.t}.\nonumber\\
\end{eqnarray}
The action then becomes

\begin{eqnarray}
S
&=& \frac{1}{2R}\int dx dy\bigg(R^2m^2\Phi^2+(S_y\partial_x\Phi+y\Delta_y\Phi)^2\nonumber\\
&+&(R^2+\frac{\kappa^2}{4})(\Delta_y\Phi)^2\bigg).
\end{eqnarray}
The classical equation of motion follows directly from this action. We get (the operators $\partial_x$ and $S_y$ commute but the order of the operators $\Delta_y$ and $y$ matters)

\begin{eqnarray}
&&(\partial_xS_y+\Delta_yy)(S_y\partial_x+y\Delta_y)\Phi+(R^2+\frac{\kappa^2}{4})\Delta_y^2\Phi\nonumber\\
&&-R^2m^2\Phi=0.\label{eom}
\end{eqnarray}
In the commutative limit we have $\kappa\longrightarrow 0$ and thus $\Delta_y=\partial_y+O(\kappa^2)$, $S_y=1+O(\kappa^2)$. As a consequence the above action reduces to

\begin{eqnarray}
S
&=& \frac{1}{2R}\int dx dy\bigg(R^2m^2\Phi^2+(\partial_x\Phi+y\partial_y\Phi)^2+R^2(\partial_y\Phi)^2\bigg).\nonumber\\
\end{eqnarray}
This action functional reads in Poincare coordinates $(t,z)$ as follows

\begin{eqnarray}
S&=& \frac{1}{2R}\int \frac{1}{R}\sqrt{g}dt dz\bigg(R^2m^2\Phi^2+z^2(\partial_z\Phi)^2+z^2(\partial_t\Phi)^2\bigg)\nonumber\\
&=& \frac{1}{2}\int \sqrt{g}d^2x\bigg(m^2\Phi^2+(\partial_a\Phi)(\partial^a\Phi)\bigg).
\end{eqnarray}
This is the correct commutative action.

In the near-boundary limit $z\longrightarrow 0$ we have  $\Delta_y=\frac{z}{R}\partial_t+O(z^3)$ and $S_y=1+O(z^2)$ since $\partial_y=\frac{z}{R}\partial_t$ (recall that $z=R\exp(-x)$ and $t=zy/R$). Thus, $S_y\partial_x+y\Delta_y=\partial_x+t\partial_t+O(z^2)=-z\partial_z+O(z^2)$. We obtain then the near-boundary action

\begin{eqnarray}
S&=& \frac{1}{2R}\int \frac{1}{R}\sqrt{g}dt dz\bigg(R^2m^2\Phi^2+z^2(\partial_z\Phi)^2\nonumber\\
&+&(1+\frac{\kappa^2}{4R^2})z^2(\partial_t\Phi)^2\bigg)\nonumber\\
&=& \frac{1}{2}\int \sqrt{g^{\prime}}d^2x^{\prime}\bigg(m^2\Phi^2+(\partial_a^{\prime}\Phi)(\partial^{\prime a}\Phi)\bigg).\label{nb}
\end{eqnarray}
The primed coordinates are given by $x^1=t^{\prime}$ and $x^2=z$ where the rescaled time parameter  $t^{\prime}$ is related to the original Poincare time $t$ by the relation

\begin{eqnarray}
t^{\prime}={(1+\frac{\kappa^2}{4R^2})^{-1/2}}t.\label{time}
\end{eqnarray}
In order to take into account the effect of boundary terms in the classical equation of motion we return to the original action (\ref{action}) and compute the variation

\begin{eqnarray}
\delta S&=&2\pi R \kappa Tr\delta\hat{\Phi}\bigg(m^2\hat{\Phi}+\frac{1}{R^2\kappa^2}[\hat{X}^a,[\hat{X}_a,\hat{\Phi}]]\bigg)\nonumber\\
&+& \pi R \kappa Tr\bigg(m^2[\hat{\Phi},\delta\hat{\Phi}]-\frac{2}{R^2\kappa^2}[\hat{X}^a,\delta\hat{\Phi}[\hat{X}_a,\hat{\Phi}]]\nonumber\\
&-&\frac{1}{R^2\kappa^2}[[\hat{X}^a,\hat{\Phi}],[\hat{X}_a,\delta\hat{\Phi}]]\bigg).\label{var}
 \end{eqnarray}
The first line produces the equation of motion of bulk fields given by  (\ref{eom}) which can then also be put in the form

\begin{eqnarray}
m^2\hat{\Phi}+\frac{1}{R^2\kappa^2}[\hat{X}^a,[\hat{X}_a,\hat{\Phi}]]=0.
 \end{eqnarray}
The second line of  (\ref{var}) represents boundary terms. These terms are all given by the trace of a commutator which should vanish identically if the trace were finite dimensional. It is straightforward to check that these boundary terms (at least up to the order of $\kappa^2$) yields the commutative result \cite{Pinzul:2017wch}

\begin{eqnarray}
-\int dt (\delta \Phi\partial_z\Phi)|_{z=0}.
 \end{eqnarray}

\section{On the ${\rm AdS}^2_{\theta}/{\rm CFT}_1$ correspondence}

The near-boundary action (\ref{nb}) can be equivalently interpreted as an action over a commutative ${\rm AdS}^2$ with Poincare coordinates $x^1=t$ and $x^2=z$ and a rescaled field $\Phi^{\prime}$ given by

\begin{eqnarray}
\Phi^{\prime}=(1+\frac{\kappa^2}{4R^2})^{1/4}\Phi.\label{nc2c}
\end{eqnarray}
The action then reads

\begin{eqnarray}
S&=& \frac{1}{2}\int \sqrt{g}d^2x\bigg((\partial_a\Phi^{\prime})(\partial^a\Phi^{\prime})+m^2\Phi^{\prime2}\bigg).\label{nb1}
\end{eqnarray}
The near-boundary action is therefore an ordinary free scalar field theory on commutative ${\rm AdS}^2$ (with rescaled fields). This means in particular that noncommutative ${\rm AdS}^2_{\theta}$ is asymptotically a commutative ${\rm AdS}^2$ (since the bounadry limit $z\longrightarrow 0$ corresponds to $x,y\longrightarrow \infty$ keeping the time $t=\exp(-x)y$ fixed).

Furthermore, it should be noted that the near-boundary action  (\ref{nb1}) is in fact exact in $\kappa$ and as such we will take it as our  first approximation of the original free scalar field theory on noncommutative ${\rm AdS}^2_{\theta}$ given by the action (\ref{action}).

The equation of motion which follows from action (\ref{nb1}) is the commutative limit of the equation of motion  (\ref{eom}). This is given explicitly by

\begin{eqnarray}
 z^2(\partial_z^2+\partial_t^2)\Phi^{\prime}-m^2R^2\Phi^{\prime}=0.\label{nb1eom}
 \end{eqnarray}
We will assume that the mass $m^2$ satisfies the so-called Breitenlohner-Freedman  bound $m^2>-1/4R^2$. The solution near the boundary $z=\epsilon$  is then given by \cite{Ramallo:2013bua}

\begin{eqnarray}
\Phi^{\prime}(t,z=\epsilon)= A(t)\epsilon^{1-\Delta}.\label{div}
\end{eqnarray}
The exponent $\Delta$ is the so-called scaling dimension of the field and it is given by

\begin{eqnarray}
\Delta=\frac{1}{2}+\sqrt{\frac{1}{4}+m^2R^2}.
\end{eqnarray}
For $m^2>0$ the exponent $1-\Delta$ is negative and hence the field (\ref{div}) is divergent. The quantum field theory source $\varphi^{\prime}(t)$, i.e. the scalar field living on the boundary, is then identified with $A(t)$, viz

\begin{eqnarray}
  \varphi^{\prime}(t)={\rm lim}_{\epsilon\longrightarrow 0}\epsilon^{\Delta-1}\Phi^{\prime}(t,\epsilon).\label{bva}
\end{eqnarray}
The field $\varphi^{'}(t)$ is the scalar field representing  the  anti-de Sitter scalar field $\Phi^{\prime}(t,z)$ at the boundary $z=0$, i.e. the boundary field $\varphi^{\prime}$ is the holographic dual of the bulk field $\Phi^{\prime}$. The scaling dimension of the source $\varphi^{\prime}$ is given by $1-\Delta$.

Let ${\cal O}^{\prime}(t,z)$ be the dual operator of the scalar field $\Phi^{\prime}(t,z)$. Their coupling is a boundary term of the form (by using also equation (\ref{nc2c}))

\begin{eqnarray}
  S_{\rm bound}&=&\int dt\sqrt{\gamma}\Phi^{\prime}(t,\epsilon){\cal O}^{\prime}(t,\epsilon)\nonumber\\
  &=&(1+\frac{\kappa^2}{4R^2})^{1/2}\int dt\sqrt{\gamma}\Phi^{}(t,\epsilon){\cal O}^{}(t,\epsilon).\label{bt0}
\end{eqnarray}
Obviously, we must have

\begin{eqnarray}
{\cal O}^{\prime}(t,z)=(1+\frac{\kappa^2}{4R^2})^{1/4}{\cal O}(t,z). \label{nc2c1}
\end{eqnarray}
It is clear that by rescaling the time coordinate in  (\ref{nb1}) we obtain the bulk action (\ref{nb}) whereas rescaling time in (\ref{bt0}) yields a canonically normalized boundary term. The time rescaling is naturally given by (\ref{time}).

The boundary term (\ref{bt0}) can also be rewritten  in terms of the dual operator ${\cal O}^{\prime}(t)$ of the scalar field $\varphi^{\prime}(t)$ as

\begin{eqnarray}
  S_{\rm bound}
  &=&\int dt \frac{R}{\epsilon} \epsilon^{1-\Delta}\varphi^{\prime}(t){\cal O}^{\prime}(t,\epsilon)\nonumber\\
  &=&R \int dt \varphi^{\prime}(t){\cal O}^{\prime}(t).\label{bt}
\end{eqnarray}
Where
\begin{eqnarray}
   {\cal O}^{\prime}(t,\epsilon)=\epsilon^{\Delta}{\cal O}^{\prime}(t).
\end{eqnarray}
This is the wave function renormalization of the operator ${\cal O}^{\prime}$ as we move into the bulk. This also shows explicitly that $\Delta$ is the scaling dimension of the dual operator ${\cal O}^{\prime}$ since going from $z=0$ to $z=\epsilon$ is a dilatation operation in the quantum field theory.

The boundary term (\ref{bt}) can be written in terms of the noncommutative scalar field $\varphi(t)$ and its dual operator ${\cal O}(t)$ as

 \begin{eqnarray}
  S_{\rm bound}
  &=&R \int dt \varphi^{}(t){\cal O}^{}(t).\label{bt1}
\end{eqnarray}
The noncommutative boundary scalar field $\varphi(t)$ and its dual operator ${\cal O}(t)$ are related to the boundary scalar field $\varphi^{\prime}(t)$ and its dual operator ${\cal O}^{\prime}(t)$ by equations (\ref{nc2c}) and (\ref{nc2c1}) respectively.

The ${\rm CFT}$ living on the boundary is completely determined by the correlation functions

 \begin{eqnarray}
\langle{\cal O}^{\prime}(t_1)...{\cal O}^{\prime}(t_n)\rangle.
 \end{eqnarray}
On the boundary with lagrangian ${\cal L}$ the calculation of these correlation functions  proceeds as usual by introducing the generating functional

 \begin{eqnarray}
Z_{\rm CFT}[J]&=&\int ~\exp({\cal L}+\int dt J(t){\cal O}^{\prime}(t))\nonumber\\
&=&\langle\exp(\int dt J(t){\cal O}^{\prime}(t))\rangle.
 \end{eqnarray}
Then we have immediately

 \begin{eqnarray}
\langle{\cal O}^{\prime}(t_1)...{\cal O}^{\prime}(t_n)\rangle=\frac{\delta^n\log Z_{CFT}[J]}{\delta J(t_1)...\delta J(t_n)}|_{J=0}.
 \end{eqnarray}
 The operator ${\cal O}^{\prime}(t)$ is sourced by the scalar field $\varphi^{\prime} (t)$ living on the boundary which is related to the boundary value of the bulk scalar field $\Phi^{\prime}(t,z)$ by the relation (\ref{bva}). The boundary scalar field $\Phi_0^{\prime}(t)$ is actually divergent and it is simply defined by

  \begin{eqnarray}
  \Phi_0^{\prime}(t)=\Phi^{\prime}(t,0).
  \end{eqnarray}
The ${\rm AdS}^2/{\rm CFT}_1$ correspondence \cite{Gubser:1998bc,Witten:1998qj} states that the ${\rm CFT}_1$ generating functional with source $J=\Phi_0^{\prime}$ is equal to the path integral on the gravity side evaluated over a bulk field which has the value $\Phi_0^{\prime}$ at the boundary of AdS. We write

  \begin{eqnarray}
Z_{\rm CFT}[\Phi_0^{\prime}]&\equiv& Z_{\rm grav}[\Phi^{\prime}\longrightarrow\Phi_0^{\prime}]\nonumber\\
&=&\int_{\Phi^{\prime}\longrightarrow \Phi_0^{\prime}} {\cal D}\Phi^{\prime}\exp(S_{\rm grav}[\Phi^{\prime}]).
 \end{eqnarray}
In the limit in which classical gravity is a good approximation the gravity path integral can be replaced by the classical amplitude given by the classical on-shell gravity action, i.e.

  \begin{eqnarray}
    Z_{\rm CFT}[\Phi_0^{\prime}]&=&\exp(S_{\rm grav}^{\rm on-shell}[\Phi^{\prime}\longrightarrow\Phi_0^{\prime}]).
 \end{eqnarray}
Typically the on-shell gravity action is divergent requiring holographic renormalization  \cite{Henningson:1998gx,deHaro:2000vlm,Skenderis:2002wp}. The on-shell action gets renormalized and the above prescription becomes

  \begin{eqnarray}
    Z_{\rm CFT}[\Phi_0^{\prime}]=\exp(S_{\rm grav}^{\rm renor}[\Phi^{\prime}\longrightarrow\Phi_0^{\prime}]).
  \end{eqnarray}
  The associated correlation functions are then renormalized as follows

  \begin{eqnarray}
\langle{\cal O}^{\prime}(t_1)...{\cal O}^{\prime}(t_n)\rangle=\frac{\delta^n S_{\rm grav}^{\rm renor}[\Phi^{\prime}\longrightarrow\Phi_0^{\prime}]}{\delta\varphi^{\prime}(t_1)...\delta\varphi^{\prime}(t_n)}|_{\varphi^{\prime}=0}.
  \end{eqnarray}
In our case the bulk action $S_{\rm grav}$ is proportional (with a proportionality constant denoted by $-\eta$) to the near-boundary action  (\ref{nb1})  which is an  approximation of the action (\ref{action}) of free scalar fields on the noncommutative ${\rm AdS}^2_{\theta}$ space.

The on-shell action $S_{\rm grav}^{\rm on-shell}$ is a boundary term obtained from $S_{\rm grav}$ by substituting the solution of the classical equation of motion (\ref{nb1eom}) (which is also required to be a regular solution in the IR limit $z\longrightarrow\infty$).

The on-shell action is found to be divergent requiring the addition of a counter term in the form of a quadratic local term living on the  boundary of ${\rm AdS}$ space given explicitly by \cite{Ramallo:2013bua}

\begin{eqnarray}
  S^{}_{\rm ct}
  &=&\frac{\eta}{2}\eta_1 \int_{} \sqrt{\gamma}dt \phi^2.
\end{eqnarray}
After some calculation we find that $\eta_1=-(1-\Delta)/R$ and as a consequence the  renormalized action reduces to the action

\begin{eqnarray}
  S^{\rm renor}_{\rm grav}
   &=&-\frac{\eta}{2}(2\Delta -1)\frac{\Gamma(1-\nu)}{\Gamma(1+\nu)}\int_{} \frac{d\omega}{2\pi} \varphi^{\prime}(\omega)(\frac{\omega}{2})^{2\nu}\varphi^{\prime}(-\omega).\nonumber\\
\end{eqnarray}
The exponent $\nu$ is given by

\begin{eqnarray}
\nu^2=\frac{1}{4}+m^2R^2=(\Delta-\frac{1}{2})^2.
\end{eqnarray}
The two-point function is then given immediately by

\begin{eqnarray}
\langle{\cal O}^{\prime}(t){\cal O}^{\prime}(0)\rangle=\frac{2\nu \eta}{\sqrt{\pi}}\frac{\Gamma(\frac{1}{2}+\nu)}{\Gamma(-\nu)}\frac{1}{|t|^{2\Delta}}.
\end{eqnarray}
This is the correct behavior of a conformal field of scaling dimension $\Delta$, i.e. the exponent $\Delta$ is indeed the scaling dimension of the boundary operator ${\cal O}^{\prime}(x)$.

The final step is to substitute the operator rescaling (\ref{nc2c1}) to obtain the noncommutative two-point function

\begin{eqnarray}
\langle{\cal O}^{}(t){\cal O}^{}(0)\rangle=(1+\frac{\kappa^2}{4R^2})^{-1/2}\frac{2\nu \eta}{\sqrt{\pi}}\frac{\Gamma(\frac{1}{2}+\nu)}{\Gamma(-\nu)}\frac{1}{|t|^{2\Delta}}.
\end{eqnarray}
By exapnding in powers of $\kappa^2$ and setting $m^2=0$ we obtain the result of \cite{Pinzul:2017wch} which was directly computed from the  noncommutative action (\ref{actionstar}) $\kappa-$expanded up to the order of $\kappa^2$.

\section{The dAFF conformal quantum mechanics}

It is a well known fact that the conformal group  in two dimensions  is an infinite dimensional symmetry group  which is generated by two copies of the Virasoro algebra (corresponding to right movers and left movers  on a closed string) and containing the group  of isometries on ${\rm AdS}^3$, i.e. containing the group ${SL}(2,R)\times {SL}(2,R)\sim SO(2,2)$ as a subgroup. There is a  ${\rm CFT}_2$, i.e. a two-dimensional conformal field theory realizing this infinite dimensional symmetry group which lives on the boundary of  ${\rm AdS}^3$.

Similarly, the conformal group in $d=1$  dimension is an infinite dimensional symmetry group generated by a single copy of the Virasoro algebra which contains as a subgroup the group  of isometries on ${\rm AdS}^2$  and which lives on the boundary of  ${\rm AdS}^2$.

Furthermore, noncommutative ${\rm AdS}^2$ is characterized by the same boundary as commutative ${\rm AdS}^2$. And, thus the commutative CFT at the boundary corresponds to the commutative ${\rm AdS}^2$ gravity theory in the bulk. However, the commutative CFT theory at the boundary in the case of  $d=1$ conformal quantum mechanics with conformal group $SO(1,2)$ is only quasi-conformal (as we will explain further shortly). Therefore, we conjecture that the noncommutative  ${\rm AdS}^2_{\theta}$ gravity theory in the bulk must be quasi-AdS for all values of the noncommutativity parameter. This has been explicitly verified for small values of $\theta$ in the previous two sections.

In this new proposal we are rather taking an inverted position, i.e. the usual position turned on its head. We are starting from the boundary and then we are moving towards the bulk and not the other way around which is the usual custom. We are thus insisting that on the boundary (of both commutative  ${\rm AdS}^2$ and noncommutative  ${\rm AdS}^2_{\theta}$) the ${\rm CFT}_1$ is  really given by conformal quantum mechanics. However, this conformal quantum mechanics is only quasi-conformal in the sense that there is neither an $SO(1,2)-$invariant quantum vacuum nor there are primary operators in the strict sense yet the bulk correlators are correctly reproduced by appropriately defined boundary correlators.

The fact that we insist that the boundary theory is given by this really quasi-conformal "conformal quantum mechanics" leads us to the conclusion that the bulk theory is not necessarily given by commutative   ${\rm AdS}^2$ which should strictly corresponds to conformal invariance. In fact,  we will argue that noncommutative  ${\rm AdS}^2_{\theta}$ is a much better candidate as it shares the same group structure as this quasi-conformal "conformal quantum mechanics" while it is quasi-AdS in the sense that it behaves more and more as commutative   ${\rm AdS}^2$ only as we approach the boundary. We are thus led to the conjecture that the conformal quantum mechanics on the boundary is actually dual to  noncommutative  ${\rm AdS}^2_{\theta}$.

At the boundary $z=0$ the isometry generators (\ref{sl2R}) reduce to the $SL(2,R)\sim SO(1,2)$ generators (with the change of notation ${\cal K}^i=K^i$)

    \begin{eqnarray}
      &&iK^1=\frac{1}{2R}(R^2+t^2)\partial_t\nonumber\\
      &&iK^2=-t\partial_t\nonumber\\
      &&iK^3=\frac{1}{2R}(R^2-t^2)\partial_t.
    \end{eqnarray}
We introduce the $su(1,1)$ generators $P$, $K$ and $D$ by the relations \cite{deAlfaro:1976vlx,Chamon:2011xk}
%\begin{eqnarray}
%  &&P=K^1-\frac{1}{2}(K^++K^-)=-iR\partial_t\nonumber\\
%  &&K=K^1+\frac{1}{2}(K^++K^-)=-i\frac{t^2\partial_t}{R}\nonumber\\
%  &&D=\frac{i}{2}(K^+-K^-)=it\partial_t\nonumber\\
%   &&D=\frac{i}{2}(K^+-K^-)=-it\partial_t.
%\end{eqnarray}
\begin{eqnarray}
  &&K=K^1+\frac{1}{2}(K^++K^-)=-i\frac{t^2\partial_t}{R}\nonumber\\
  &&P=K^1-\frac{1}{2}(K^++K^-)=-iR\partial_t\nonumber\\
%  &&D=\frac{i}{2}(K^+-K^-)=it\partial_t\nonumber\\
   &&D=-\frac{i}{2}(K^+-K^-)=+it\partial_t.
\end{eqnarray}
The $so(1,2)$ Lie algebra (\ref{so12})  becomes
%\begin{eqnarray}
%  i[D,H]=H~,~i[D,K]=-K~,~i[K,H]=2D.
%\end{eqnarray}
\begin{eqnarray}
  i[D,P]=P~,~i[D,K]=-K~,~i[K,P]=2D.\label{SL2R}
\end{eqnarray}
This $SL(2,R)$ structure can be extended to the infinite dimensional conformal group in one dimenions by the relation $L_n=i\hat{t}^{n+1}\partial_{\hat{t}}$ where $\hat{t}=-t/R$. Of course, $L_0=D$ (scale transformations or dilatations), $L_{+1}=K$ (special conformal transformations) and $L_{-1}=P$ (translations). We have then the commutation relations
\begin{eqnarray}
    [L_m,L_n]=-i(m-n)L_{m+n}.
\end{eqnarray}
This is the Lie algebra of  the infinite dimensional conformal group in $d=1$  dimension. A careful derivation of this Lie algebra in the context of  ${\rm AdS}^2$ and its theories can be found for example in \cite{Cadoni:1998sg,Cadoni:1999ja}. Thus, $P\equiv L_{-1}$ generates translations, $D\equiv L_0$ generates scale transformations and $K=L_{+1}$ generates  special conformal transformations. A general  ${SL}(2,R)$ transformation is given by
\begin{eqnarray}
\hat{t}\longrightarrow \hat{t}^{\prime}=\frac{\alpha \hat{t}+\beta}{\gamma \hat{t}+\delta}~,~\alpha\delta -\beta\gamma=1.
\end{eqnarray}
%We will also need the operator $R$ defined by
%\begin{eqnarray}
%  R=K^1=\frac{1}{2}(K+P).
%\end{eqnarray}
The Casimir $C$ of the $SL(2,R)$ group takes then the form
\begin{eqnarray}
 -C=(K^1)^2-K^1-K^+K^-=\frac{1}{2}(PK+KP)-D^2.\label{cas}\nonumber\\
\end{eqnarray}
The eigenstates and eigenvalues in the lowest weight representations $D_k^+$ are alternatively denoted by
\begin{eqnarray}
 |km\rangle\equiv |n\rangle~,~m\equiv r_0+n~,~r_0\equiv k>0.\label{st1}
\end{eqnarray}
In the irreducible representation  $D_k^+$ the "magnetic" quantum number $n$ takes the integer values $n=0,1,2,...$ and the "negative" Casimir operator takes the value $-C=r_0(r_0-1)$ where the "pseudo-spin" quantum number $r_0$ is assumed to only take the integer values $r_0=1,2,...$.

The dAFF conformal quantum mechanics, which is a quantum model covariant under the $SL(2,R)$ algebra (\ref{SL2R}), was constructed originally in $1976$ by de Alfaro, Fubini and Furlan \cite{deAlfaro:1976vlx}.

The action of the generators $D$, $P$ and $K$ on a scalar field $\Phi$ is given by the equations
\begin{eqnarray}
 [D,\Phi(t)]=-i\big(t\partial_t\Phi(t)+\Delta \Phi(t)\big).\label{h1}
\end{eqnarray}
\begin{eqnarray}
 [P,\Phi(t)]=-i\partial_t\Phi(t).\label{h2}
\end{eqnarray}
\begin{eqnarray}
 [K,\Phi(t)]=-i(t^2\partial_t+2\Delta t)\Phi(t).\label{h3}
\end{eqnarray}
The number $\Delta$ is called the scaling dimension of the scalar field $\Phi$. This scalar field furnishes therefore an irreducible representation of the conformal algebra (\ref{SL2R}) which is characterized by the scaling dimension $\Delta$. The commutators (\ref{h1}), (\ref{h2}) and (\ref{h3}) yield the transformation law  under the scale transformations $t\longrightarrow t^{\prime}=\lambda t$  
\begin{eqnarray}
  \Phi(t)\longrightarrow \Phi^{\prime}(t^{\prime})=\lambda^{-\Delta}\Phi(t).\label{tl}
\end{eqnarray}
In conformal field theory the scaling dimension (which is the eigenvalue of the dilatation operator) plays the role of the energy and as a consequence it is the dilatation operator $D$ which plays the role of the (conformal) Hamiltonian with time parameter given by $\tau=\ln t$. This can be seen as follows.

In Euclidean spacetime with $d\geq 2$ which is foliated using spheres $\mathbb{S}^{d-1}$ it is observed that the action of the dilatation operator allows us to move from one sphere to another (radial quantization). The metric reads (with $\Omega$ the solid angle on $\mathbb{S}^{d-1}$ and $\tau=\log r$)
\begin{eqnarray}
ds^2&=&dr^2+r^2d\Omega^2\nonumber\\
&=&e^{2\tau}(d\tau^2+d\Omega^2).
\end{eqnarray}
This metric is conformally equivalent to the metric on the cylinder. In other words, the transformation $r\longrightarrow \tau=\log r$ maps $\mathbb{R}\times \mathbb{S}^{d-1}$ to $\mathbb{R}^{d}$. The parameter $\tau$ plays then the role of the time parameter. The lower base of the cylinder is at the infinite past $\tau\longrightarrow -\infty$ ($r=0$) whereas the upper base of the cylinder is at the infinite future $\tau\longrightarrow +\infty$ ($r=+\infty$). Going around the cylinder is given by the solid angle    $\Omega$. The evolution operator is then given by 
\begin{eqnarray}
U|\Delta\rangle=\exp(i\tau D)|\Delta\rangle.
\end{eqnarray}
In conformal quantum mechanics (our case) with $d=1$ this result remains valid with the identification $\tau=\ln t$ and hence scale transformations $t\longrightarrow t^{\prime}=\lambda t$ are seen as (conformal) time translations $\tau\longrightarrow \tau^{\prime}=\tau+\ln\lambda$. The transformation law (\ref{tl}) takes then the quantum mechanical form
\begin{eqnarray}
  \Phi^{\prime}(t)=U^{\dagger}_{\lambda}\Phi(t)U_{\lambda}~,~U_{\lambda}=e^{i\ln\lambda D}.
  \end{eqnarray}
For infinitesimal transformations we obtain

\begin{eqnarray}
  \Phi^{\prime}(t)=\Phi(t)-i\ln\lambda[D,\Phi(t)]+....
  \end{eqnarray}
However, the infinitesimal form of the transformation law (\ref{tl}) is actually of the form
\begin{eqnarray}
  \Phi^{\prime}(t)=\Phi(t)-\ln\lambda\big(t\partial_t\Phi(t)+\Delta \Phi(t)\big)+....
\end{eqnarray}
By comparing these two last equations we get the action of the dilatation operator $D$ on fields $\Phi$ with scaling dimensions $\Delta$ given by equation (\ref{h1}). Similar calculations yield the action of the momentum operator $P$ and the action of the special conformal generator $K$ on scalar fields $\Phi$ given by equations (\ref{h2}) and (\ref{h3}) respectively.

We also note that the momentum operator $P$ generates translations and acts as a raising operator while the special conformal generator $K$ generates  special conformal transformations and acts as a lowering operator.

An operator annihilated by  $K$ is called a primary operator. By acting on this primary operator with  $P$ we obtain the so-called descendant operators. The primary operator and its descendant operators form a conformal family.

An  irreducible representation of the conformal algebra (\ref{SL2R}) characterized by the scaling dimension $\Delta$ is therefore furnished by a primary operator ${\cal O}_{\Delta}(0)$ satisfying 
\begin{eqnarray}
[D,{\cal O}_{\Delta}(0)]=-i\Delta {\cal O}_{\Delta}(0).\label{pr1}
\end{eqnarray}

\begin{eqnarray}
[P,{\cal O}_{\Delta}(0)]=-i\dot{{\cal O}_{\Delta}}(0).\label{pr3}
\end{eqnarray}

\begin{eqnarray}
[K,{\cal O}_{\Delta}(0)]=0.\label{pr2}
\end{eqnarray}
In conformal field theory with $d\geq 2$ there is a unique vacuum state $|\Omega\rangle$ which is invariant under the global conformal group $SO(1,d)$. We will assume for a moment that such a state exists also in $d=1$ dimensions, viz
\begin{eqnarray}
  D|\Omega\rangle=P|\Omega\rangle=K|\Omega\rangle=0.\label{st2}
  \end{eqnarray}
Then it is not difficult to verify that the state $|\Delta\rangle={\cal O}_{\Delta}(0)|\Omega\rangle$ is a highest weight state, also called primary state, satisfying
\begin{eqnarray}
  D|\Delta\rangle=-i\Delta|\Delta\rangle~,~ K|\Delta\rangle=0.
  \end{eqnarray}
This is the primary state at the origin $t=0$ corresponding to the primary operator  ${\cal O}_{\Delta}(0)$. By inserting the operator  ${\cal O}_{\Delta}(t)$ at an arbitrary point $t$ we will create the state $|\chi\rangle={\cal O}_{\Delta}(t)|\Omega\rangle$. All other states (descendant states) can be obtained by acting successively with the raising operator $P$.

By using equations (\ref{pr1}), (\ref{pr3}), (\ref{pr2}) which define primary operators and equations (\ref{st2}) which defines the ground state $|\Omega\rangle$ we can compute the boundary two-point and three-point functions and find them in full agreement with the bulk ${\rm AdS}^2$ two-point and three-point functions. This statement/fact is the central content of the ${\rm AdS}^2/{\rm CFT}_1$ correspondence. 

However the situation is much more involved in one dimension.

Let us first recall what happens in higher dimensions. In conformal field theory with $d\geq 2$ there is a unique vacuum state $|0\rangle$ which is invariant under the global conformal group $SO(d,2)$. This corresponds to no operator insertion in the cylinder which would create a state at a given time $\tau$ (corresponding to a given radius $r$). Let ${\cal O}_{\Delta}(x)$ be some operator with scaling dimension $\Delta$. The insertion of this operator at the origin $r=0$ (or infinite past $\tau=-\infty$) creates the state $|\Delta\rangle={\cal O}_{\Delta}(0)|0\rangle$ with scaling dimension $\Delta$. By inserting the operator ${\cal O}_{\Delta}(x)$ at an arbitrary point $x$ will create the state 
 \begin{eqnarray}
|\chi\rangle={\cal O}_{\Delta}(x)|0\rangle&=&\exp(iPx) {\cal O}_{\Delta}(0)\exp(-iPx)|0\rangle\nonumber\\
&=&\exp(iPx) |\Delta\rangle.
\end{eqnarray}
This is a linear superposition of states with different eigenvalues $\Delta$ since the momentum operator $P_{\mu}$ define raising operators with respect to the eigenvalues of the dilatation operator, i.e. it raises the scaling dimension $\Delta$ by unity. From the other hand, the special conformal generator  $K_{\mu}$ defines lowering operators with respect to the eigenvalues of the dilatation operator, i.e. it lowers the scaling dimension $\Delta$ by unity.

An operator annihilated by  $K_{\mu}$ is called a primary operator. By acting on this primary operator with  $P_{\mu}$ we obtain the so-called descendant operators. The primary operator and its descendant operators form a conformal family.

An infinite dimensional irreducible representation of the conformal group is determined by an irreducible representation of the Lorentz group with definite conformal dimension and annihilated by the special conformal generator $K_{\mu}$. The stability algebra at the origin consists of the generators $D$, $K_{\mu}$ (and the Lorentz generators $M_{\mu\nu}$). A primary conformal operator ${\cal O}_{\Delta}$ (the lowest weight state) in a given representation of the Lorentz group is defined by
\begin{eqnarray}
[P_{\mu},{\cal O}_{\Delta}(0)]=-i\partial_{\mu}{\cal O}_{\Delta}(0).
\end{eqnarray}
\begin{eqnarray}
[D,{\cal O}_{\Delta}(0)]=-i\Delta {\cal O}_{\Delta}(0).
\end{eqnarray}
\begin{eqnarray}
[K,{\cal O}_{\Delta}(0)]=0.
\end{eqnarray}
The descendants $\partial...\partial {\cal O}(0)$ are obtained by the repeated action of the momentum operators $P_{\mu}$. The eigenvalues of the Lorentz operators $M_{\mu\nu}$ on the primary operator  ${\cal O}$ are spin quantum numbers denoted for example by $j_L$ and $j_R$. This defines an irreducible representation of the conformal group characterized by $\Delta$, $j_L$ and $j_R$.

In conformal quantum mechanics with $d=1$ and conformal group $SO(1,2)$ much of these results remain valid with few crucial exceptions. First, we note that the Lorentz  generators $M_{\mu\nu}$ and as a consequence the  spin quantum numbers $j_L$ and $j_R$ are absent.

The most serious discrepancy between  conformal quantum mechanics in $d=1$ with conformal group $SO(1,2)$ and between conformal field theory in $d\geq 2$ with conformal group $SO(1,d)$ is the absence in conformal quantum mechanics of a normalized $SO(1,2)-$invariant vacuum state $|\Omega\rangle$ which is annihilated by all the generators, i.e a vacuum state which satisfies (\ref{st2}) does not or can not exist. The main reason behind this is the fact that quantum mechanics requires a single-particle Hilbert space whereas quantum field theory requires a Fock space which is the direct sum of tensor products of single-particle Hilbert spaces corresponding to different number of particles including a genuinely empty Hilbert space  (the vacuum state) \cite{Chamon:2011xk}.

The other difference between conformal quantum mechanics and conformal field theory is the absence of primary operators satisfying (\ref{pr1}), (\ref{pr3}) and (\ref{pr2}). However, by using these equations which define primary operators and equations (\ref{st2}) which defines the ground state $|\Omega\rangle$ we obtain the condition (for any real number $x$)
\begin{eqnarray}
  (xK+iD){\cal O}_{\Delta}(0)|\Omega\rangle=\Delta {\cal O}_{\Delta}(0)|\Omega\rangle.\label{st000}
\end{eqnarray}
This condition with $x=1/2$, as we will show next, is sufficient to reproduce bulk ${\rm AdS}^2$ correlators by appropriate boundary operators which are not strictly speaking primary operators \cite{Chamon:2011xk}. 

First, we introduce a coherent-like state $|t\rangle$ associated with the time variable $t$ on the bondary by the Schrodinger equations of motion, associated with the Heisenberg equations of motion (\ref{h1}), (\ref{h2}) and (\ref{h3}) respectibvely, given by %\footnote{Remember that there is a sign difference between the Heisenberg picture ($[H,A]=-i\frac{d}{dt}A$) and the Schrodinger picture ($H|\psi\rangle=i\frac{d}{dt}|\psi\rangle$). }
\begin{eqnarray}
 D|t\rangle=-i\big(t\partial_t+\Delta \big)|t\rangle.\label{h111}
\end{eqnarray}
\begin{eqnarray}
 P|t\rangle=-i\partial_t|t\rangle.\label{h222}
\end{eqnarray}
\begin{eqnarray}
 K|t\rangle=-i(t^2\partial_t+ 2\Delta t)|t\rangle.\label{h333}
\end{eqnarray}
We can check that the Casimir operator in this $t-$representation takes the value $C=\Delta(\Delta-1)$ which means that the scaling dimension of the state $|t\rangle$ is $\Delta=r_0$.  We expand the states $|t\rangle$ in the states $|n\rangle$, given in equation (\ref{st1}), then we solve the equations of motion (\ref{h111}), (\ref{h222}) and (\ref{h333}) for $|t\rangle$. Let us then start by writing
\begin{eqnarray}
 |t\rangle=\sum_n\langle n|t\rangle |n\rangle=\sum_n\beta_n^*(t)|n\rangle.
\end{eqnarray}
We compute the states  $|t\rangle$ explicitly by computing the function $\beta_n(t)$. We start from $\langle t|K^1|n\rangle=r_n\langle t|n\rangle$ where we recall that $|n\rangle=|km\rangle$ and $r_n=r_0+n=m$. But $2\langle t|K^1|n\rangle=\langle t|(K+P)|n\rangle=i(t^2\partial_t+2r_0t+\partial_t)\langle t|n\rangle$ \footnote{The operators $P=p^2/2+g/q^2$ (with $g=3/8+2r_0(r_0-1)$), $D=tP-(qp+pq)/4$ and $K=-t^2P+2tD+q^2/2$ are hermitian.}. Thus, we obtain the differential equation 
\begin{eqnarray}
 \frac{i}{2}\bigg[(t^2+1)\frac{d}{dt}+2r_0t\bigg]\beta_n(t)=r_n\beta_n(t).
\end{eqnarray}
The solution is of the form
\begin{eqnarray}
\beta_n(t)=\beta_n^{(1)}(t)\bigg(\frac{1-it}{1+it}\bigg)^{r_n}.
\end{eqnarray}
The remainder $\beta_n^{(1)}(t)$ satisfies the homogeneous equation 
\begin{eqnarray}
 \frac{i}{2}\bigg[(t^2+1)\frac{d}{dt}+2r_0t\bigg]\beta_n^{(1)}(t)=0.
\end{eqnarray}
This can be solved by $\beta_n^{(1)}(t)=v_n/(1+t^2)^{r_0}$ with $v_n$ being a constant and hence we obtain the solution
\begin{eqnarray}
\beta_n(t)=\frac{v_n}{(1+t^2)^{r_0}}\bigg(\frac{1-it}{1+it}\bigg)^{r_n}.\label{beta}
\end{eqnarray}
The normalization $v_n$ is given by \cite{Chamon:2011xk}
\begin{eqnarray}
  v_n=(-1)^n\sqrt{\frac{\Gamma(2r_0+n-1)}{n!}}.\label{beta1}
\end{eqnarray}
This will be verified shortly. The states $|n\rangle$ can be computed in the usual way using $K^{\pm}|n\rangle=\sqrt{r_n(r_n\pm 1)-r_0(r_0-1)}|n\pm 1\rangle$. We should set $|n\rangle=\alpha_n(K^+)^n|0\rangle$. We can then check for example using $[K^1,(K^+)^n]=n(K^+)^n$ that $K^1|n\rangle=r_n|n\rangle$ as it should be. From the normalization condition $\langle n|n\rangle=1$ and the repeated use of $K^-|n\rangle=\sqrt{n(2r_0+n-1)}|n-1\rangle$ we compute 
\begin{eqnarray}
  1&=&\alpha_n^*\langle 0|(K^-)^n|n\rangle\nonumber\\
  &=&\alpha_n^*\sqrt{n!\frac{\Gamma(2r_0+n-1)}{\Gamma(2r_0-1)}}\Rightarrow \alpha_n=\sqrt{\frac{\Gamma(2r_0-1)}{n!\Gamma(2r_0+n-1)}}.\label{beta2}\nonumber\\
\end{eqnarray}
We can now recheck directly the two relations  $K^-|n\rangle=\sqrt{n(2r_0+n-1)}|n-1\rangle$ and $ K^+|n\rangle=\sqrt{(n+1)(2r_0+n)}|n+1\rangle$.

By substituting $\beta_n(t)$ given by (\ref{beta}) in $|t\rangle$ and using the explicit values of $v_n$ and $\alpha_n$ given by (\ref{beta1}) and (\ref{beta2}) we obtain the solution
\begin{eqnarray}
  |t\rangle&=&\bigg(\frac{\omega+1}{2}\bigg)^{2r_0}\sqrt{\Gamma(2r_0-1)}\sum_{n=0}\frac{(-1)^n}{n!}(\omega K^+)^n|0\rangle\nonumber\\
  &=&N(t)\exp(-\omega K^+)|0\rangle.
  \end{eqnarray}
The new variable $\omega$ is given in terms of the time variable $t$ by the relation $\omega=(1+it)/(1-it)$ whereas the normalization $N(t)$ is given by
\begin{eqnarray}
  N(t)&=&\bigg(\frac{\omega+1}{2}\bigg)^{2r_0}\sqrt{\Gamma(2r_0-1)}.\label{beta3}
  \end{eqnarray}
We can check explicitly that $|t\rangle$ is the correct solution of say (\ref{h111}) as follows. We start from the definition
\begin{eqnarray}
  D|t\rangle&=&-\frac{i}{2}(K^+-K^-)|t\rangle\nonumber\\
  &=&-\frac{i}{2}N(t)\bigg(-\frac{d}{d\omega}e^{-\omega K^+}-[K^{-},e^{-\omega K^+}]\bigg)|0\rangle.\nonumber\\
\end{eqnarray}
By using $[K^-,(K^+)^n]=n(n-1)(K^+)^{n-1}+2n(K^+)^{n-1}K^1$ we can show that $[K^-,e^{-\omega K^+}]=e^{-\omega K^+}(\omega^2K^+-2\omega K^1)$. It is now straightforward to show that
\begin{eqnarray}
  D|t\rangle
  &=&-ir_0\omega|t\rangle-\frac{i}{2}(1-\omega^2)\frac{d\ln N}{d\omega}|t\rangle+\frac{i}{2}(1-\omega^2)\frac{d}{d\omega}|t\rangle\nonumber\\
  &=&-ir_0\omega|t\rangle-\frac{i}{2}(1-\omega^2)\frac{d\ln N}{d\omega}|t\rangle-it\frac{d}{dt}|t\rangle.
\end{eqnarray}
But from (\ref{h111}) we must have
\begin{eqnarray}
  D|t\rangle
   &=&-ir_0|t\rangle-it\frac{d}{dt}|t\rangle.
\end{eqnarray}
In other words, we must have $d\ln N/d\ln (\omega+1)=2r_0$. This confirms the normalization (\ref{beta3}) and as a consequence the normalization (\ref{beta1}).

In summary, we have obtained  \cite{Chamon:2011xk}
\begin{eqnarray}
  |t\rangle&=&O(t)|0\rangle\nonumber\\
  &=&N(t)\exp(-\omega K^+)|0\rangle.\label{Ot}
  \end{eqnarray}
The new variable $\omega$ and the normalization $N(t)$ are given by
\begin{eqnarray}
  \omega=\frac{1+it}{1-it}~,~N(t)&=&\bigg(\frac{\omega+1}{2}\bigg)^{2r_0}\sqrt{\Gamma(2r_0-1)}.\label{beta3}\nonumber\\
  \end{eqnarray}
The state $|t\rangle$ is a coherent-like state in the sense that it is an  eigenstate of $K^{-}+\omega K^1$ with eigenvalue $-r_0\omega$. Indeed, we compute
\begin{eqnarray}
  (K^-+\omega K^1)|t\rangle
   &=&-r_0\omega|t\rangle.\label{virtue}
\end{eqnarray}

\section{The operator-state correspondence}
In order to construct the operator-state correspondence we need in principle to construct an invariant vacuum state together with primary states and primary operators. A primary operator ${\cal O}_{\Delta}(t)$ with scaling dimension $\Delta$ is defined by the conditions
\begin{eqnarray}
&&[P,{\cal O}_{\Delta}(0)]=-i\dot{{\cal O}_{\Delta}}(0)\nonumber\\
&&[D,{\cal O}_{\Delta}(0)]=-i\Delta {\cal O}_{\Delta}(0)\nonumber\\
&&[K,{\cal O}_{\Delta}(0)]=0.\label{one}
\end{eqnarray}
The corresponding primary state $|{\cal O}_{\Delta}(t)\rangle={\cal O}_{\Delta}(t)|\Omega\rangle$ is constructed from this primary operator  ${\cal O}_{\Delta}(t)$ and from a vacuum state $|\Omega\rangle$ which is $SO(2,1)-$invariant, i.e. it is annihilated by all the generators, viz 
\begin{eqnarray}
  D|\Omega\rangle=P|\Omega\rangle=K|\Omega\rangle=0.\label{two}
\end{eqnarray}
In this section we will extensively employ implicitly and explicitly the coherent-like/temporal-like basis $\{|t\rangle\}$. We can immediately compute

\begin{eqnarray}
  \bigg(\frac{1}{2}K+iD\bigg)e^{-K^+}&=&\bigg(\frac{1}{2}K^1+\frac{3}{4}K^+-\frac{1}{4}K^-\bigg)e^{-K^+}\nonumber\\
  %&=&e^{-K^+}\bigg(\frac{1}{2}K^1+\frac{3}{4}K^+-\frac{1}{4}K^-\bigg)\nonumber\\
  %&+&\frac{1}{2}[K^1,e^{-K^+}]-\frac{1}{4}[K^-,e^{-K^+}]\nonumber\\
  &=&e^{-K^+}\bigg(\frac{1}{2}K^1+\frac{3}{4}K^+-\frac{1}{4}K^-\bigg)\nonumber\\
  &-&\frac{1}{2}e^{-K^+}K^+-\frac{1}{4}e^{-K^+}\bigg(K^+-2K^1\bigg)\nonumber\\
  &=&e^{-K^+}(K^1-\frac{1}{4}K^-).
\end{eqnarray}
This means in particular that we must have
\begin{eqnarray}
  \bigg(\frac{1}{2}K+iD\bigg)O(0)|0\rangle&=&r_0O(0)|0\rangle.\label{four}
\end{eqnarray}
We choose then $|{\cal O}_{\Delta}(0)\rangle=|t=0\rangle=O(0)|0\rangle$. Furthermore, we compute
\begin{eqnarray}
  K^1e^{-P}|{\cal O}_{\Delta}(0)\rangle%&=&\bigg(e^{-P}R+[R,e^{-P}]\bigg)|{\cal O}_{\Delta}(0)\rangle\nonumber\\
  &=&e^{-P}\bigg(K^1+iD-\frac{1}{2}P\bigg)|{\cal O}_{\Delta}(0)\rangle\nonumber\\
  &=&e^{-P}\bigg(\frac{1}{2}K+iD\bigg)|{\cal O}_{\Delta}(0)\rangle\nonumber\\
  &=&r_0e^{-P}|{\cal O}_{\Delta}(0)\rangle.\label{three1}
\end{eqnarray}
This last result shows explicitly that the state $e^{-P}|{\cal O}_{\Delta}(0)\rangle$ is an eigenstate of $K^1$ with eigenvalue $\Delta$, e.g. $e^{-P}|{\cal O}_{\Delta}(0)\rangle\propto |0\rangle$ with $\Delta=r_0$.

Thus, although the vacuum state $|0\rangle$ is not an $SO(1,2)-$invariant state (it does not satisfy the condition (\ref{st2})) and the operators $O(t)$ are not strictly speaking primary operators (they do not satisfy the conditions (\ref{pr1}), (\ref{pr3}), (\ref{pr2})) we observe that equations (\ref{four}) is precisely equation (\ref{st000}) (with $x=1/2$ and $\Delta=r_0$).

In other words, the operators $O(t)$  and the corresponding states $|t\rangle=O(t)|0\rangle$, despite all shortcoming, behave effectively as the primary operators ${\cal O}_{\Delta}(t)$ and the primary states $|{\cal O}_{\Delta}(t)\rangle={\cal O}_{\Delta}(t)|\Omega\rangle$ with scaling dimension $\Delta=r_0$. We have then in the dAFF conformal quantum mechanics the quasi-vacuum state, the quasi-primary operators and the quasi-primary states given respectively by 
\begin{eqnarray}
&&|0\rangle\longrightarrow |\Omega\rangle\nonumber\\
&&O(t)=N(t)e^{-\omega K^+}\longrightarrow  {\cal O}_{\Delta}(t)~,~\Delta=r_0\nonumber\\
&&|t\rangle=O(t)|0\rangle\longrightarrow |{\cal O}_{\Delta}(t)\rangle={\cal O}_{\Delta}(t)|\Omega\rangle.
\end{eqnarray}
This conclusion can also be reached by calculating explicitly the two-point and three-point correlation functions following \cite{deAlfaro:1976vlx,Chamon:2011xk}. We find (with $B(t)$ a primary operator with scaling dimension $\Delta=\delta$) the results
\begin{eqnarray}
  \langle 0|O^{\dagger}(t_1)O(t_2)|0\rangle=\frac{f_0}{(t_1-t_2)^{2r_0}}.\label{fu1}
\end{eqnarray}
\begin{eqnarray}
  \langle 0|O^{\dagger}(t_1)B(t)O(t_2)|0\rangle=\frac{f_0}{(t-t_1)^{\delta}(t-t_2)^{\delta}(t_1-t_2)^{2r_0-\delta}}.\nonumber\\\label{fu2}
\end{eqnarray}
These boundary correlation functions are precisely the bulk correlation functions obtained in the commutative ${\rm AdS}^2$.

Let us give a brief demonstration of these two fundamental results (\ref{fu1}) and (\ref{fu2}). The two-point correlation function is defined by 
\begin{eqnarray}
  F_2(t_1,t_2)&=&\langle 0|O^{\dagger}(t_1)O(t_2)|0\rangle\nonumber\\
  &=&\langle t_1|t_2\rangle\nonumber\\
  &=&\sum_n\beta_n(t_1)\beta_n^*(t_2).
\end{eqnarray}
This can be calculated indirectly, allowing us to show explicitly the underlying conformal symmetric origin of the result, as follows. From the action of the $SO(2,1)-$generators $P$, $D$ and $K$ on the coherent-like state $|t\rangle$ given by equations (\ref{h111}), (\ref{h222}) and (\ref{h333}) we can show that the correlation function $F(t_1,t_2)$ must satisfy the differential equations
\begin{eqnarray}
   &&(\partial_{t_1}+\partial_{t_2})F_2(t_1,t_2)=0\nonumber\\
  &&(t_1\partial_{t_1}+t_2\partial_{t_2}+2r_0)F_2(t_1,t_2)=0\nonumber\\
   &&(t_1^2\partial_{t_1}+t_2^2\partial_{t_2}+2r_0t_1+2r_0t_2)F_2(t_1,t_2)=0.
\end{eqnarray}
The first equation leads immediately the result $F_2(t_1,t_2)=f(t)$ where $t=t_1-t_2$. By replacing the ansatz $f(t)=f_0t^{\delta}$ in the second or third equation we obtain the exponent $\delta=-2r_0$. Hence the two-point function is given by
\begin{eqnarray}
  F_2(t_1,t_2)&=&\frac{f_0}{(t_1-t_2)^{2r_0}}. \label{F2}
\end{eqnarray}
Next, we compute the three-point function 
\begin{eqnarray}
  F_3(t_1,t,t_2)&=&\langle 0|O^{\dagger}(t_1)B(t)O(t_2)|0\rangle\nonumber\\
  &=&\langle t_1|B(t)|t_2\rangle\nonumber\\
  &=&\sum_{n_1,n_2}\beta_{n_1}(t_1)\beta_{n_2}^*(t_2)\langle n_1|B(t)|n_2\rangle.\nonumber\\
\end{eqnarray}
The operator $B(t)$ is a primary operator with scaling dimension $\Delta=\delta$, viz

\begin{eqnarray}
&& [D,B(t)]=-i\big(t\partial_tB(t)+\delta B(t)\big)\nonumber\\
&& [P,B(t)]=-i\partial_tB(t)\nonumber\\
&& [K,B(t)]=-i(t^2\partial_tB(t)+2\delta tB(t)).\label{ze}
\end{eqnarray}
We compare $\langle t_1|[X,B(t)]|t_2\rangle$ (computed using equations (\ref{ze})) with $\langle t_1|XB(t)|t_2\rangle-\langle t_1|B(t)X|t_2\rangle$ (computed using   equations (\ref{h111}), (\ref{h222}) and (\ref{h333})) to derive the differential equations
\begin{eqnarray}
   &&(\partial_{t_1}+\partial_{t_2}+\partial_t)F_3=0\nonumber\\
  &&(t_1\partial_{t_1}+t_2\partial_{t_2}+t\partial_t+2r_0+\delta)F_3=0\nonumber\\
   &&(t_1^2\partial_{t_1}+t_2^2\partial_{t_2}+t^2\partial_t+2r_0t_1+2r_0t_2+2\delta t)F_3=0.\nonumber\\
\end{eqnarray}
From the first equation we conclude that $F_3$ must be of the form $F_3=f(t-t_1,t-t_2,t_1-t_2)$. We consider the ansatz $f(t)=f_0(t-t_1)^{A_1}(t-t_2)^{A_2}(t_1-t_2)^{A_3}$. The second equation gives the constraint $A_1+A_2+A_3+2r_0+\delta=0$. The third equation gives the constraints $A_1+A_3=-2r_0$, $A_1+A_2=-2\delta$ and $A_2+A_3=-2r_0$. The solution is found to be given by $A_1=A_2=-\delta$ and $A_3=\delta-2r_0$. The three-point function is then given by
\begin{eqnarray}
  F_3(t_1,t,t_2)&=&\frac{f_0}{(t-t_1)^{\delta}(t-t_2)^{\delta}(t_1-t_2)^{2r_0-\delta}}. \label{F3}
\end{eqnarray}
The boundary correlation functions given by equations (\ref{F2}) and (\ref{F3}) are precisely the bulk correlation functions obtained in the commutative ${\rm AdS}^2$. Since the same boundary is common to both commutative  ${\rm AdS}^2$ and noncommutative  ${\rm AdS}^2_{\theta}$ this boundary result is expected to hold also for  noncommutative  ${\rm AdS}^2_{\theta}$. This is another, more precise, meaning of the statement that noncommutative  ${\rm AdS}^2_{\theta}$ is a quasi-AdS space in the same sense that the dAFF conformal quantum mechanics is really only a quasi-conformal theory.

\section{The noncommutative geometry of the commutative boundary}

In analogy with the "north pole of the sphere" we define the "north pole of ${\rm AdS}^2$" by $(X_1,X_2,X_3)=(R,0,0)$. Around this point (defined in terms of $SO(1,2)$ coherent states) noncommutative ${\rm AdS}^2_{\theta}$ will appear as a noncommutative Moyal-Weyl plane $\mathbb{R}^2_{\theta}$ (flattening limit and  stereographic projection). The algebra of operators (\ref{polten}) reduces then to the algebra of operators at the boundary given explicitly by

\begin{eqnarray}
&&T_{KM}=f_{KM}(\hat{X}^+)^M~,~M\geq 0\nonumber\\
&&T_{KM}=f_{KM}(\hat{X}^-)^{-M}~,~M\leq 0.\label{polten1}
\end{eqnarray}
$f_{KM}$ are now constants. The algebra of quasi-primary operators $O(0)$ at the boundary given by  (\ref{Ot}) is precisely generated by the algebra of operators (\ref{polten1}). Indeed, it is seen that the operators $O(0)$ given by  (\ref{Ot}) can be expanded in terms of the polarization tensors $T_{KM}$ with positive $M$ evaluated at the north pole  of ${\rm AdS}^2_{\theta}$. The algebra (\ref{polten1}) defines actually the noncommutative geometry (spectral triple) of the commutative boundary.

The action of the Laplacian (\ref{Laplacian}) on the operators (\ref{polten1}) or equivalently on the quasi-primary operators $O(0)$ reduces to the action of the outer derivation ${\cal K}^1$ by virtue of the relation (\ref{virtue}) which when evaluated at $t=0$ reads $K^-+K^1=-r_0$. We have then the one-dimensional Laplacian

\begin{eqnarray}
{\cal K}^2&=&-{\cal K}_1^2+{\cal K}_2^2+{\cal K}_3^2\nonumber\\
&=&-{\cal K}^1({\cal K}^1+1)+{\cal K}^-{\cal K}^+\nonumber\\&\equiv& -{\cal K}^1({\cal K}^1+1).\label{Laplacian1}
\end{eqnarray}
In a similar vein the action of the conformal generators $D$, $P$ and $K$ on the  quasi-primary operators $O(0)$ reduce to the action of the outer derivation ${\cal K}^1$, viz
\begin{eqnarray}
&&[D,O(0)]=-\frac{i}{2}[K^1,O(0)]\nonumber\\
&&[P,O(0)]=\frac{3}{2}[K^1,O(0)]\nonumber\\
&&[K,O(0)]=\frac{1}{2}[K^1,O(0)].
\end{eqnarray}
And
\begin{eqnarray}
[K^1,O(0)]=(K^++2K^1)O(0).
\end{eqnarray}
This shows explicitly why the operators $O(t)$ are quasi-primary operators. And also it shows that the outer derivation ${\cal K}^1$ is effectively the only independent derivation on the boundary.

The spectral triple defining the one-dimensional boundary is then a subalgebra of the spectral triple defining noncommutative ${\rm AdS}^2_{\theta}$. It consists of the algebra (\ref{polten1}) together with the Laplacian (\ref{Laplacian1}) whereas the corresponding Hilbert space is implicitly defined by the Hilbert space ${\cal H}_k^{\pm}$ of  noncommutative ${\rm AdS}^2_{\theta}$.

This construction of the boundary is reminiscent of the construction of the fuzzy circle $\mathbb{S}^1_N$ from the spectral triple of the fuzzy sphere $\mathbb{S}^2_N$ \cite{Dolan:2003kq}.

\section{A Moyal-Weyl bulk-boundary map}

The $SL(2,R)$ algebra (\ref{SL2R}) satisfied by the  $so(2,1)=su(1,1)$ Lie algebra generators $P$, $D$ and $K$ can also be realized in terms of a canonical pair $(q,p)$ satisfying the canonical Heisenberg algebra 
\begin{eqnarray}
  [\hat{q},\hat{p}]=i\hbar.
\end{eqnarray}
As a consequence this pair $(\hat{q},\hat{p})$ which is defined on the boundary should be mapped to the canonical coordinate operators $(\hat{x},\hat{y})$ defined in the bulk of noncommutative ${\rm AdS}^2_{\theta}$ and satisfy the Heisenberg algebra (\ref{MW}). The operator map is an isomorphism between the two corresponding Hilbert spaces and it is given explicitly by 

\begin{eqnarray}
&&\hat{q}\equiv\sqrt{\frac{\hbar}{\kappa}}\hat{x}=\sqrt{\frac{\hbar}{\kappa}}\ln\big(\frac{\hat{X}_1-\hat{X}_3}{R}\big)\nonumber\\
&&\hat{p}\equiv\sqrt{\frac{\hbar}{\kappa}}\hat{y}=\sqrt{\frac{\hbar}{\kappa}}\hat{X}_2.\label{identification}
\end{eqnarray}
The commutative limit $\kappa\longrightarrow 0$ of the noncommutative geometry of the bulk must then be correlated with the classical limit $\hbar\longrightarrow 0$ of the quantum mechanics on the boundary in such a way that $\kappa/\hbar$ is kept fixed while $(\hat{q},\hat{p})$ approaches the classical phase space $(q,p)$ and $(\hat{x},\hat{y})$ approaches the commutative ${\rm AdS}^2$.

The noncommutative  ${\rm AdS}^2_{\theta}$ is a quasi-AdS space, i.e. it looks more and more like a commutative ${\rm AdS}^2$ as we approach the boundary. Thus, near the boundary the operators $\hat{x}$ and $\hat{y}$ approach the commutative coordinates $x$ and $y$ respectively. In fact, the two boundaries are approached when $x\longrightarrow \infty$ and $y\longrightarrow \pm \infty$ with $z=R\exp(-x)\longrightarrow 0$ and $t=\exp(-x)y$ kept fixed.

Strictly speaking the two boundaries are approached at large and almost-common eigenvalues of  the operators $\hat{x}$ and $\hat{y}$. Hence from the above identification (\ref{identification}) it is seen that the quantum mechanical operators $\hat{q}$ and $\hat{p}$  living on the boundary contain information about both the bulk and the boundary of noncommutative ${\rm AdS}^2_{\theta}$. The large and almost-common eigenvalues of  $\hat{q}$ and $\hat{p}$ really captures the geometry of the boundary whereas the geometry of the bulk is captured by the eigenvalues located away from these limits.

An alternative interpretation goes as follows. By solving the Heisenberg equations of motion on the boundary the quantum mechanical operators  $\hat{q}$ and $\hat{p}$ are found to be functions of the single coordinate operator $\hat{t}= \exp(-\hat{x})\hat{y}$ (recall that $\hat{x}$ and $\hat{y}$ almost-commute near the boundary). We define now the operators  $\hat{q}$ and $\hat{p}$ in the bulk by means of the identification (\ref{identification}). These two definitions together define the Moyal-Weyl bulk-boundary operator map between the canonical coordinate operators $(\hat{x},\hat{y})$ of noncommutative ${\rm AdS}^2_{\theta}$ and the quantum mechanical operators  $(\hat{q},\hat{p})$ on the boundary.

In the dAFF conformal quantum mechanics the $so(2,1)=su(1,1)$ Lie algebra generators $P$, $D$ and $K$ are realized in terms of a single degree of freedom $q(t)$  with conjugate momentum $p(t)=\dot{q}(t)$ satisfying the Heisenberg algebra
\begin{eqnarray}
  [q,p]=i.
\end{eqnarray}
We start from their $SO(2,1)$ action given by the equations
\begin{eqnarray}
  [P,q]=-i\dot{q}\Rightarrow P=\frac{p^2}{2}+V(q).
\end{eqnarray}
And
\begin{eqnarray}
  [D,q]&=&-it\dot{q}-i\Delta q\nonumber\\
  &=&t[P,q]+\frac{\Delta}{2}[pq+qp,q]\nonumber\\
  &&\Rightarrow D=tP+\frac{\Delta}{2}(qp+pq)+V_1(q).
\end{eqnarray}
In this equation $\Delta$ is the scaling dimension of the operator $q$. And the remaining $SO(2,1)$ action is given by the equation
\begin{eqnarray}
  [K,q]&=&-it^2\dot{q}-2i\Delta t q\nonumber\\
  &=&t^2[P,q]+2t([D,q]+it\dot{q})\nonumber\\
  &=&-t^2[P,q]+2t[D,q]\Rightarrow K=-t^2P+2tD+V_2(q).\nonumber\\
\end{eqnarray}
We must also have the Lie algebra
\begin{eqnarray}
[D,P]=-iP~,~[D,K]=iK~,~[K,P]=-2iD.\nonumber\\
\end{eqnarray}
From the first equation we compute the commutator
\begin{eqnarray}
[D,P]=i\Delta p^2-i\Delta q V^{\prime}(q)+\frac{i}{2}(V_1^{\prime}(q)p+pV_1^{\prime}(q)).
\end{eqnarray}
But this must be equal to $-iP$, i.e. 
\begin{eqnarray}
  [D,P]=-i\frac{p^2}{2}-iV(q).
\end{eqnarray}
By comparing we necessarily get  (with $g>0$)
\begin{eqnarray}
\Delta=-\frac{1}{2}~,~V(q)=\frac{g}{q^2}~,~V_1(q)=0.
\end{eqnarray}
Next we compute the commutator
\begin{eqnarray}
  [D,K]=it^2P-\frac{it}{2}(V_2^{\prime}(q)p+pV_2^{\prime}(q))+\frac{i}{2}qV_2^{\prime}(q).
\end{eqnarray}
But this must be equal to $iK$, i.e. 
\begin{eqnarray}
  [D,K]=it^2P-\frac{it}{2}(qp+pq)+iV_2^{}(q).
\end{eqnarray}
By comparing we further get the requirement 
\begin{eqnarray}
V_2(q)=\frac{1}{2}q^2.
\end{eqnarray}
The final commutator $[K,P]=-2iD$ checks out trivially. By using now the results $P=p^2/2+g/q^2$, $D=tP-\{q,p\}/4$ and $K=-t^2P+2tD+q^2/2$ we compute the Casimir operator

\begin{eqnarray}
  -C&=&\frac{1}{2}(PK+KP)-D^2\nonumber\\
  &=&\frac{1}{4}\{P,q^2\}-\frac{1}{16}\{q,p\}^2\nonumber\\
  &=&\frac{g}{2}+\frac{1}{8}\{p^2,q^2\}-\frac{1}{16}\{q,p\}^2\nonumber\\
  &=&\frac{g}{2}-\frac{3}{16}\nonumber\\
  &\equiv& r_0(r_0-1).
\end{eqnarray}
The degree of freedom $q\equiv q(t)$ satisfies the Heisenberg equations of motion (\ref{h1}), (\ref{h2}) and (\ref{h3}) with scaling dimension $\Delta=-1/2$, viz

\begin{eqnarray}
 [D,q]=-i\big(t\dot{q}-\frac{1}{2} q\big).\label{h11}
\end{eqnarray}
\begin{eqnarray}
 [P,q]=-i\dot{q}.\label{h22}
\end{eqnarray}
\begin{eqnarray}
 [K,q]=-i(t^2\dot{q}- t q).\label{h33}
\end{eqnarray}

\section{Conclusion}

This article contains the first part of our study in which we attempt a coherent unification between the principles of noncommutative geometry (and their matrix models) from the one hand and the principles of the ${\rm AdS}^{2}/{\rm CFT}_1$ correspondence from the other hand. In this part the main focus has been on constructing a consistent ${\rm QM}/{\rm NCG}$ correspondence, i.e. a duality between:
\begin{itemize}
\item The dAFF conformal quantum mechanics (${\rm QM}$) on the boundary (which is really only "quasi-conformal" in the sense of field theory) on the one hand.
\item And from the other hand the noncommutative geometry of ${\rm AdS}^2_{\theta}$  (${\rm NCG}$) in the bulk (which is a "quasi-AdS" space in the sense of being only asymptotically ${\rm AdS}^2$).
\end{itemize}
Noncommutative geometry is understood here as "first quantization" of geometry whereas the corresponding Yang-Mills matrix models provide "quantum gravity" or "second quantization" of the corresponding geometry. Thus, noncommutative ${\rm AdS}^2_{\theta}$ is the first quantization of commutative ${\rm AdS}^2$ whereas quantum-gravitational fluctuations around ${\rm AdS}^2_{\theta}$ are captured by the gauge/gravitational fluctuations of the Yang-Mills IKKT-type matrix models.

The symmetry structure given here by the Lorentz group $SO(1,2)$ is the starting point of the noncommutative geometry but it also  the unifying structure underlying: 1) the ${\rm AdS}^2$ spacetime, 2) the noncommutative ${\rm AdS}^2_{\theta}$ space,  3) the ${\rm CFT}_1$ theory on the boundary given by the dAFF conformal quantum mechanics, 4) the Yang-Mills IKKT-type matrix models, and 5) the geometry of the boundary  (which is common to both commutative ${\rm AdS}^2$ and noncommutative ${\rm AdS}^2_{\theta}$).

The logic followed in this article in constructing the  ${\rm QM}/{\rm NCG}$ goes as follows:
\begin{itemize}

\item In this novel proposal the customary understanding of the ${\rm AdS}^2/{\rm CFT}_1$ correspondence is turned on its head. It is customary to assumed that the   boundary theory in the case  of ${\rm AdS}^2$ is essentially unkown while the bulk theory is given by some gravity theory about commutative ${\rm AdS}^2$. Here instead we assume that the boundary theory is completely known given by the dAFF conformal quantum mechanics then we seek an appropriate gravity theory residing in the bulk.

\item The intrinsic difficulty in the case  of ${\rm AdS}^2$  is then traced to the fact that ${\rm CFT}_1$ as given by conformal quantum mechanics is really  only "quasi-conformal" and as a consequence the gravity theory in the bulk is only required to be quasi-AdS.

\item Thus, in this proposal we are turning our understanding of the ${\rm AdS}^{2}/{\rm CFT}_1$ on its head since we are starting from the boundary and then we are moving towards the bulk and not the other way around. We are thus insisting that on the boundary the ${\rm CFT}_1$ is  really given by dAFF conformal quantum mechanics. However, this conformal quantum mechanics is only "quasi-conformal" in the sense that there is neither an $SO(1,2)-$invariant quantum vacuum state nor there are primary operators in the strict sense yet the bulk correlators are correctly reproduced by appropriately defined boundary quantum fields.

\item The fact that we insist that the boundary theory is given by this quasi-conformal "conformal quantum mechanics" leads us to the conclusion that the bulk theory is not necessarily given by commutative   ${\rm AdS}^2$ which strictly corresponds to conformal invariance. In fact,  it is argued that noncommutative  ${\rm AdS}^2_{\theta}$ is a much better candidate as it shares the same group structure as this quasi-conformal "conformal quantum mechanics" while it is quasi-AdS in the sense that it behaves more and more as commutative   ${\rm AdS}^2$ as we approach the boundary.

\item It is further observed that the Lorentz group $SO(1,2)$ is the fundamental unifying structure of the ${\rm AdS}^2$ spacetime, the noncommutative ${\rm AdS}^2_{\theta}$ space and of the boundary quantum theory. In particular, the algebra of quasi-primary operators at the boundary is seen to be a subalgebra of the operator algebra of  noncommutative ${\rm AdS}^2_{\theta}$. This leads us to the conclusion/conjecture that the theory in the bulk must be given by noncommutative geometry and not by classical gravity, i.e. it is given by noncommutative ${\rm AdS}^2_{\theta}$ and not by commutative ${\rm AdS}^2$. Thus, the "quasi-conformal" dAFF conformal quantum mechanics on the boundary is actually dual to the "quasi-AdS"  noncommutative  ${\rm AdS}^2_{\theta}$ in the bulk.

\end{itemize}

\appendix
\section{Brief remark on Lorentzian ${\rm AdS}^2_{\theta}$}
The quantization of  Lorentzian ${\rm AdS}^2_{\theta}$ goes essentially through the same steps. The defining relations are
\begin{eqnarray}
  -\hat{X}_1^2-\hat{X}_2^2+\hat{X}_{3}^2=-R^2.\label{well1L}
\end{eqnarray}
%This is a noncommutative space in which the coordinate operators $\hat{X}^a$ are found to satisfy the commutation relations

\begin{eqnarray}
  [\hat{X}^a,\hat{X}^b]=-i\kappa \epsilon^{ab}~_c\hat{X}^c.\label{well2L}
\end{eqnarray}
The coordinate operators $\hat{X}^a$ which solve (\ref{well1L}) and (\ref{well2L}) are still given by (\ref{well3}) but now $K^a$ are the generators of the Lie group $SO(2,1)$ in the irreducible representations of the Lie algebra $[K^a,K^b]=-i\epsilon^{ab}~_cK^c$  given by the continuous series $C_k^{\frac{1}{2}}$.  Indeed, for Lorentzian ${\rm AdS}^2_{\theta}$ the Casimir in the discrete and finite representations is positive whereas in the continuous and complementary series the Casimir is negative. This selects the continuous and complementary representations but the complementary is not admissible since there is no large $k$ limit.

Thus, the relation between the deformation parameter $\kappa$ and the $su(1,1)$ spin quantum number $j\equiv k-1$ is still given by (\ref{rel1}) and as a consequence the commutative limit is again given by (\ref{rel2}).

However, the space of operators (noncommutative functions) on the Lorentzian ${\rm AdS}^2_{\theta}$ is quite different from the algebra (\ref{pr1}) of operators on Euclidean ${\rm AdS}^2_{\theta}$. Indeed, noncommutative functions on the Lorentzian ${\rm AdS}^2_{\theta}$ are given by the following tensor product
\begin{eqnarray}
C_k^{\epsilon}\otimes C_{k^{'}}^{\epsilon^{'}}=\bigoplus_{K=K_{\rm min}}^{\infty}D_K^{+}\oplus\bigoplus_{K=K_{\rm min}}^{\infty}D_K^{-}\oplus 2 \int_{\mathbb{R}_+}^{\bigoplus}C^{E}_{\frac{1}{2}+is} ds.\nonumber\\
\end{eqnarray}
If $\epsilon+\epsilon^{'}$ is an integer than $K_{\rm min}=E=0$ whereas if $\epsilon+\epsilon^{'}$ is half-integer then $K_{\rm min}=E=1/2$. In the current case $k=k^{'}$, $\epsilon=\epsilon^{'}=1/2$, $K_{\rm min}=E=0$.

These noncommutative functions form an orthonormal basis with the scalar product (\ref{sp}) and they are in fact $SU(1,1)$ polarization tensors $T_{KM}$ satisfying

\begin{eqnarray}
{\cal K}^2T_{KM}=K(K-1)T_{KM}~,~{\cal K}^3T_{KM}=MT_{KM}.
\end{eqnarray}
It is not difficult to verify again the behavior  \cite{Ho:2000fy,Ho:2000br,Jurman:2013ota}
\begin{eqnarray}
&&T_{KM}=f_{KM}(\hat{X}^3)(\hat{X}^+)^M~,~M\geq 0\nonumber\\
&&T_{KM}=f_{KM}(\hat{X}^3)(\hat{X}^-)^{-M}~,~M\leq 0.
\end{eqnarray}
This is precisely the correct behavior of commutative functions on the commutative Lorentzian ${\rm AdS}^2$.

\end{document}